\documentclass[a4paper,12pt,oneside]{amsart}
\usepackage{graphicx}
\usepackage{amssymb}
\usepackage{paralist}
\usepackage{cite}
\usepackage{amscd}
\usepackage{dsfont}
\usepackage{bm}

\newtheorem{theorem}{Theorem}[section]
\newtheorem{lemma}[theorem]{Lemma}
\newtheorem{prop}[theorem]{Proposition}
\newtheorem{rem}[theorem]{Remark}   
\newtheorem{rems}[theorem]{Remarks}   
\newtheorem{col}[theorem]{Corollary }  

\theoremstyle{definition}

\theoremstyle{remark}

\numberwithin{equation}{section}

\newcommand{\diag}[1]{\textrm{diag}\set{#1}}

\newcommand{\set}[1]{\left\{#1\right\}}

\def \1{\mathds{1}}
\def \0{\mathsf{0}}

\def\pclone{\stackrel{p}{\sim}}
\def\ppclone{\stackrel{p+1}{\sim}}

\def\P{\mathcal{P}}
\def\Cll{ \mathcal{C}}
\def\Cl{C}
\def\Z{ \mathcal{Z}}
\def\N{ \mathcal{N}}
\def\Q{ {\mathcal{Q}}}
\def\T{ {\mathcal{T}}}
\def\X{ {\mathcal{X}}}

\def\q{ {\mathsf{q}}}
\def\p{ {\mathsf{p}}}

\def\F{ {\mathcal{F}}}

\def\H{ {\mathcal{H}}}

\def\trace{ {\mathrm{Tr}}}

\def\Re{ {\mathrm{Re}}}

\begin{document}

\title[Counting of Sieber-Richter pairs]{Counting of Sieber-Richter pairs of periodic orbits}

\author{Boris Gutkin${}^\dag$, Vladimir Osipov${}^*$}
\address{
${}^\dag$ Faculty of Physics, University Duisburg-Essen, Lotharstr. 1, 47048 Duisburg, Germany;
\newline ${}^*$ Institute of Theoretical Physics, Cologne University
Zülpicher Str. 77, 50937 Cologne, Germany.
}
\email{boris.gutkin@uni-duisburg-essen.de}
\begin{abstract}
In the framework of the semiclassical approach the universal spectral correlations in the Hamiltonian systems with classical chaotic dynamics can be attributed to the systematic correlations between actions of periodic orbits which (up to the switch in the momentum direction)  pass through approximately the same points of the phase space. By considering symbolic dynamics of the system one can  introduce a natural ultrametric  distance between  periodic orbits and organize them into clusters. Each cluster consists of orbits approaching closely each other in the phase space.  We  study the distribution of cluster sizes for the backer's map in the asymptotic limit of long trajectories. This problem is equivalent to the one  of counting degeneracies in the  length spectrum  of  the {\it de Bruijn} graphs. Based on this fact, we derive the  probability $\P_k$ that $k$ randomly chosen periodic orbits belong to the same cluster.  Furthermore, we find  asymptotic behaviour of the largest cluster size $|\Cll_{\max}|$ and derive the probability $P(t)$ that a random periodic orbit belongs to a cluster of the size smaller than $t|\Cll_{\max}|$, $t\in[0,1]$. 
\end{abstract}

\maketitle

\section{Introduction}

\begin{figure}[htb]
\begin{center}{
a)\includegraphics[height=3.0cm]{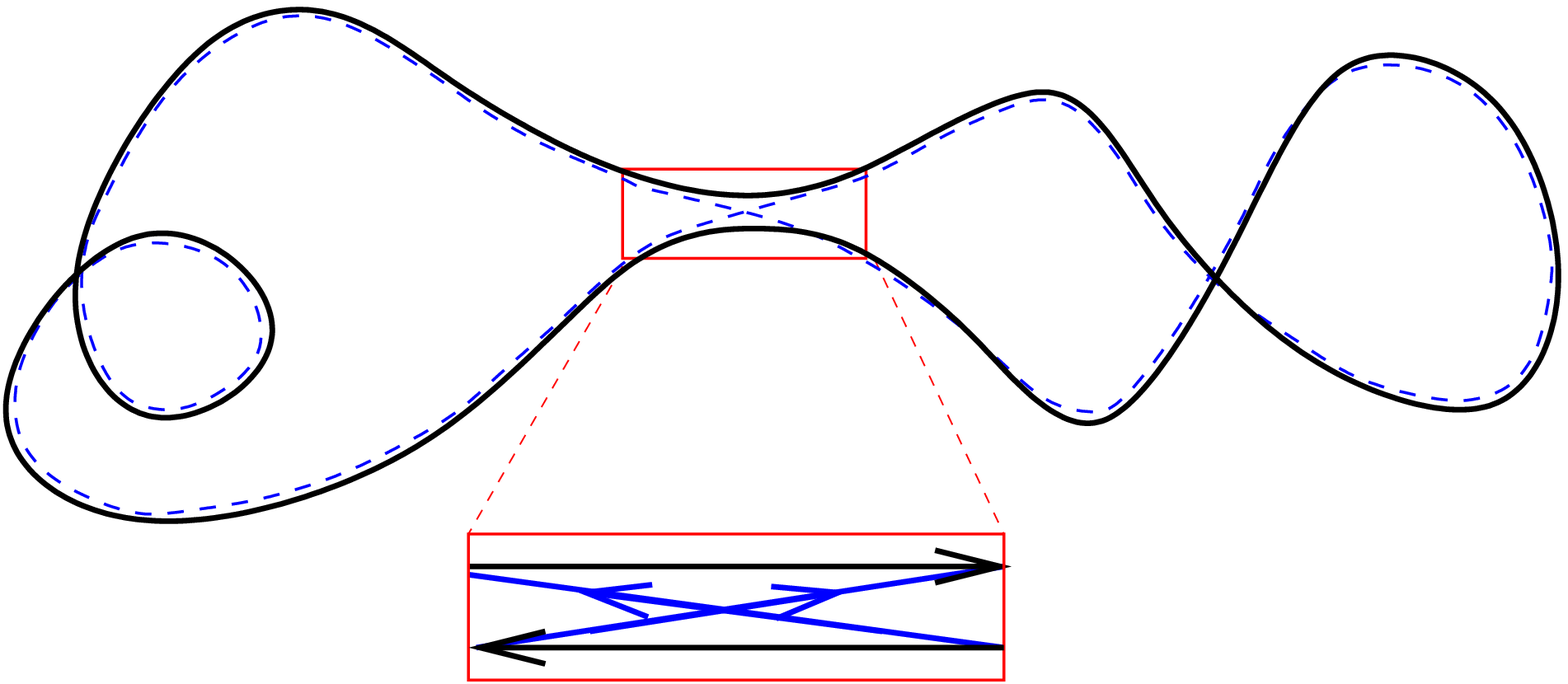}\hskip 0.5cm b)\includegraphics[height=2.2cm]{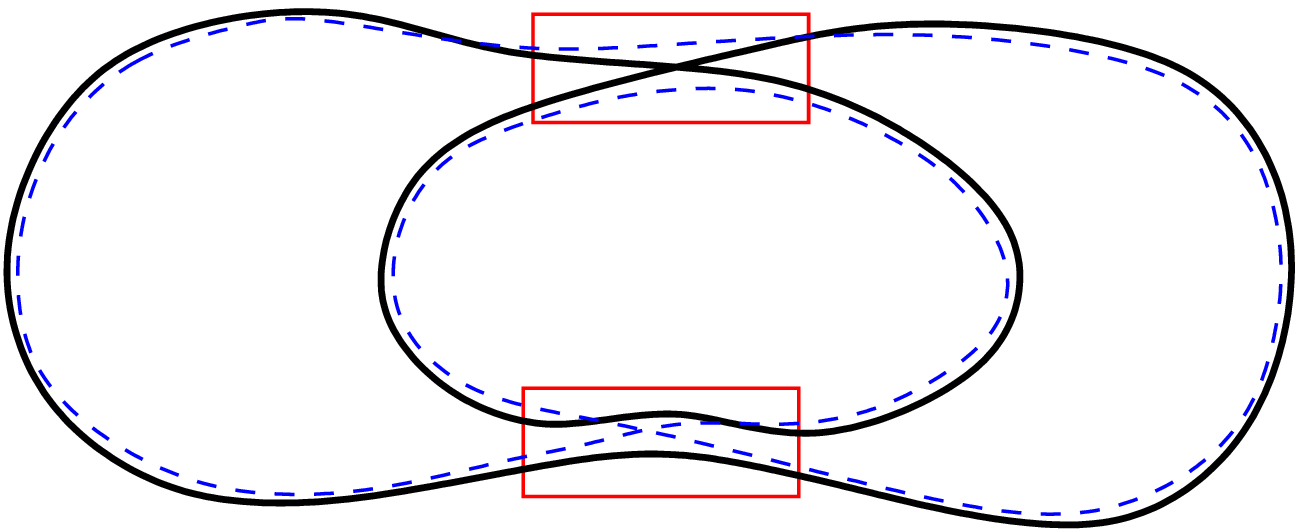}
}\end{center}
\caption{ \small{a) Schematic picture of Sieber-Richter pair of periodic orbits with one encounter region being marked out by a box. b) Pair of periodic orbits with two encounters. As opposed to the case (a), here two orbits pass close to each other  in the  phase space of the system. Only such orbits contribute to the spectral correlations of systems with broken  time-reversal symmetry \cite{haake1}.  }  }\label{sieber_richter}
\end{figure}

In their seminal paper \cite{bgs} Bohigas Giannoni and Schmit conjectured that the local energy spectrum statistics of quantum systems with fully chaotic classical dynamics  are universal and can  be described by standard ensembles of Random Matrix Theory (RMT). To explore the origin of such universality semiclassical techniques based on the applications of the  Gutzwiller  trace formula was introduced by Berry in~\cite{berry}. By this approach  the correlations between energy levels of a  quantum Hamiltonian system can be related to the  correlations between periodic orbit actions  in the corresponding classical system. Within the diagonal approximation, where only correlations between periodic orbits themselves are taken into account, Berry managed to obtain the leading order of the universal spectral form factor. The diagonal approximation, however, turned out to be  insufficient  to reproduce full RMT result, whose derivation  had remained  a distinguished challenge for yet a long time~\cite{ADDKKSS, haake}. The breakthrough was achieved in 2001 when Sieber and Richter discovered  a non-trivial mechanism of correlations between periodic orbit actions~\cite{sr}. They showed that the next to the leading order term of the universal spectral form factor can be obtained by taking into account correlations  between long periodic orbits with one self-crossing   under a small angle (usually referred as {\it encounter})  and the ass!
 ociated partner orbits,  see fig.~\ref{sieber_richter}a. Such pairs of orbits have close actions and contribute systematically into the spectral correlations.   Later, this approach was extended to include correlations between periodic orbits having an arbitrary number of encounters which culminated in the derivation of the full RMT result \cite{haake1}. 

In a nutshell Sieber-Richter pairs and its many-encounter analogs  are nothing more then bunches of  periodic orbits running through almost the same  points  of the phase space up to the switch in the momentum direction. Furthermore, in the case of broken time reversal symmetry  only orbits   passing close to each other with the same  momentum direction  are of relevance, see fig.~\ref{sieber_richter}b.  In the present paper we restrict  consideration only to  the latter case. Due to the hyperbolic nature of the dynamics the action difference between such trajectories  is small and determined by the lengths of the encounters. Equally important, the correlation mechanism   of Sieber-Richter pairs is  robust to perturbations of the dynamical system. In fact, any hyperbolic system contains a large number of long periodic orbits with close actions which are not of Sieber-Richter type. However, it might be expected that, in general, the differences  between  their actions fluctuate enormously under perturbations of the system. Therefore,  the contribution from a generic pair of periodic orbits     is washed out  after averaging (e.g., over ensemble of  systems)  and only  pairs  of Sieber-Richter type contribute systematically to the spectral correlations. 

The robustness of   Sieber-Richter correlation mechanism can be easily understood by considering symbolic dynamics of the system. Assuming that a finite Markov partition exists, any periodic orbit can be encoded by a finite sequence of symbols $x=x_1 x_2\dots x_n $ from some alphabet \cite{licht}. The fact that two periodic orbits  come close to each other in the phase space has a natural interpretation on the level of the corresponding symbol sequences. Namely,
let $x$ and  $y$ be two sequences of  length $n$ such that any subsequence of  length $p$ occurs in $x$ exactly the same number of times as in $y$. This property of sequences $x$ and $y$ is referred below as {\it $p$-closeness}.  It is straightforward  to see that  any pair of  $p$-close
sequences  defines in fact a   pair of   periodic orbits which are close with respect to the Euclidean metric  of  the phase space. Their metric closeness  is controlled by $p$:  the larger  parameter $p$ is, the closer  two periodic orbits approach each other. 
Accordingly, all periodic orbits of the system can be organized into a disjoint union of {\it clusters} of $p$-close orbits. Each cluster is composed of periodic orbits  with  close actions  running through approximately the same points of the phase space in a different time order.

Motivated by the application to quantum chaos, questions regarding  the number  of periodic trajectories with close actions/lengths were previously addressed both in  physics and mathematics literature in the context of several different models of chaotic dynamical systems:  geodesics flows on manifolds of negative curvature \cite{sharp1}, billiards \cite{uzy3,petkov}, quantum maps \cite{uzy1} and quantum graphs \cite{ uzy2,sharp2,tanner,Berkolajko}. In the present paper we study the phenomenon of clustering of periodic orbits    on the level of symbolic dynamics. Our consideration is restricted to the simplest possible grammar assuming a two-letter alphabet, $x_i\in\{0,1\}$, and absence of pruning, i.e. each symbol in the sequence  can be followed by any other symbol.  These grammar rules is met, for instance, in the  baker's map~\cite{backer}.  The main question to be addressed below can be informally stated  as follows: given an integer $p$, what is the probability that a randomly picked up periodic sequence (equiv. orbit)  of length $n$ has a certain number of $p$-close partners when $n$ is large enough?  

\section{Definitions and main results}
\subsection{The baker's map} In what follows  we  consider clustering  of periodic orbits within the paradigm model of chaotic system -- the baker's map $\T$.   Explicitly, the action of   $\T$  on the points $v=(\q,\p)$ of the two-dimensional phase space $V=[0,1)\times [0,1)$ is given by:
\begin{equation}
 \T\cdot v=\begin{cases}
           (2\q,\frac{1}{2}\p)& \text{if $\q\in[0,\frac{1}{2})$},\\
           (2\q-2,1-\frac{1}{2}\p)& \text{if $\q\in[\frac{1}{2}, 1)$},
          \end{cases}
\end{equation}
where $\q$ and $\p$ play the role of the coordinate and momentum respectively, see e.g., \cite{backer} for details. The
 baker's map has an advantage of having a particularly simple symbolic dynamics. This allows to  avoid cumbersome notation and  makes the exposition more transparent.

\subsection{Symbolic dynamics and periodic orbits} Symbolic dynamics is a standard tool widely used in the theory of hyperbolic dynamical systems. By this approach  each point  of the system phase space is identified with a sequence of symbols from a certain alphabet. Given such a representation the time evolution of the system takes a   simple form.
  
To  introduce symbolic  dynamics it is necessary first to define  Markov partition of the phase space $V$. A standard choice for the baker's map is  $V_0\sqcup V_1$, where $V_0=[0,\frac{1}{2})\times [0,1)$ and $V_1=[\frac{1}{2}, 1)\times [0,1)$. Then any point $v=(\q,\p)\in V$ can be uniquely encoded by a two-sided sequence $x_-.x_+$ of zeros and ones: $x_+=x_1x_2\dots$, $x_- =\dots x_{-2}x_{-1}x_{0}$, $x_i\in\{0,1\}$ using a simple algorithm: $x_i=0$ if $\T^i v\in V_0$ and  $x_i=1$ if $\T^i v\in V_1$ for $i\in\mathbb{Z}$. Positive $i$ correspond to the ``future'' evolution of $v$, which is written in the $x_+$ subsequence. The coordinate $q$ expresses through this subsequence by the formula $ \q=0.x_+$. The ``past'' history of  $v$ and the momentum $\p$ are defined by $x_-$ subsequence: $ \p=0.x_-$.

Within the above symbolic representation the time evolution of the system  is given by the shift map 
\[\sigma:\; [\dots x_{-1}x_{0}.x_1x_2 x_3\dots]\to[\dots x_{-1}x_{0}x_1.x_2 x_3\dots],\]
 which  moves  the  separation point ``$.$'' between the future and the past in the sequence of symbols step by step. 

All infinite periodic sequences composed of one and the same finite piece $x\in X_n$ correspond to periodic orbits of the system. Here and below symbol $X_n$ stands for the set of all possible sequences of zeroes and ones having the length $n$. Let $\gamma_x$ denotes the periodic orbit of the backer's map associated with the sequence $x\in X_n$. Note, that two sequences $x$, $x'$ correspond to the same periodic trajectory if and only if they are related by the cyclic shift, i.e. if $x=x_1 x_2 \dots x_n$ and $x'=x_{i+1}\dots x_n x_1\dots  x_i$ for some $i\in\{1,\dots n-1\}$ then correspond to one and the same periodic trajectory $\gamma_x=\gamma_{x'}$. In what follows we will also consider the quotient set $\X_n:=X_n/\sim$ with respect to the cyclic shift $x\sim x'$.  It is convenient to think of the elements belonging to the set $\X_n$ as of sequences from $X_n$ with the ``glued'' ends.  Importantly, the elements from $\X_n$ and all periodic orbits having the period $n$ are in one to one correspondence according to the remark made above in this paragraph.

\subsection{Clusters of periodic orbits} To define the  clusters of  periodic orbit, firsts,  we need to  introduce the notion of their closeness.  Take two $n$-periodic orbits $\gamma_x$, $\gamma_y$  composed of $n$ points  $\{\gamma_x(i)\}_{i=0}^{n-1}$, $\{\gamma_y(i)\}_{i=0}^{n-1}$ in the phase space. It is natural to think about two orbits $\gamma_x$,  $\gamma_y$  as of close ones if in a vicinity of any point  $\gamma_x(i)$, $i=0,\dots n-1$ one can find some point from the set $\{\gamma_y(i)\}_{i=0}^{n-1}$ and vice versa. In other words, two trajectories pass through approximately the same parts of the phase space but perhaps in a different order.  To put this picture on a more solid background, we say $\gamma_x$ is in the $p$-neighborhood  of  $\gamma_y$ if there exist exactly $n$ pairs $\left(\gamma_x(i_k), \gamma_y(i'_k) \right)$, $k=1,\dots n$ such that for each $k$ distance between points is bounded by  $||\gamma_x(i_k), \gamma_y(i'_k)||\leq 2^{-p}$, where the distance $||v,v'||$ in the phase space between $v=(\q,\p)$ and $v'= (\q',\p')$ is defined as $||v,v'||:=\max\{|\q-\q'|, |\p-\p'|\} $.

The above notion of metric closeness between periodic trajectories can be carried over to the topological space $\X_n$.  We will say that two sequences $x,y\in \X_n$ are $p$-close if any sequence of $p\leq n$ consecutive symbols $a_1 a_2 \dots a_p$, $a_i \in\{0,1\}$  appears the same number of times (which might be also zero) both in  $x$ and  $y$. Speaking informally $x$ and  $y$ are $p$-close if their local content (of the length $p$) is exactly the same in both sequences.  This equivalence relation is denoted below as $x \pclone   y$. It is  straightforward to see that  $\gamma_x$ is in the $p$-neighborhood  of  $\gamma_y$  whenever $x \pclone  y$.

\begin{figure}[htb]
\begin{center}{
\includegraphics[height=7.8cm]{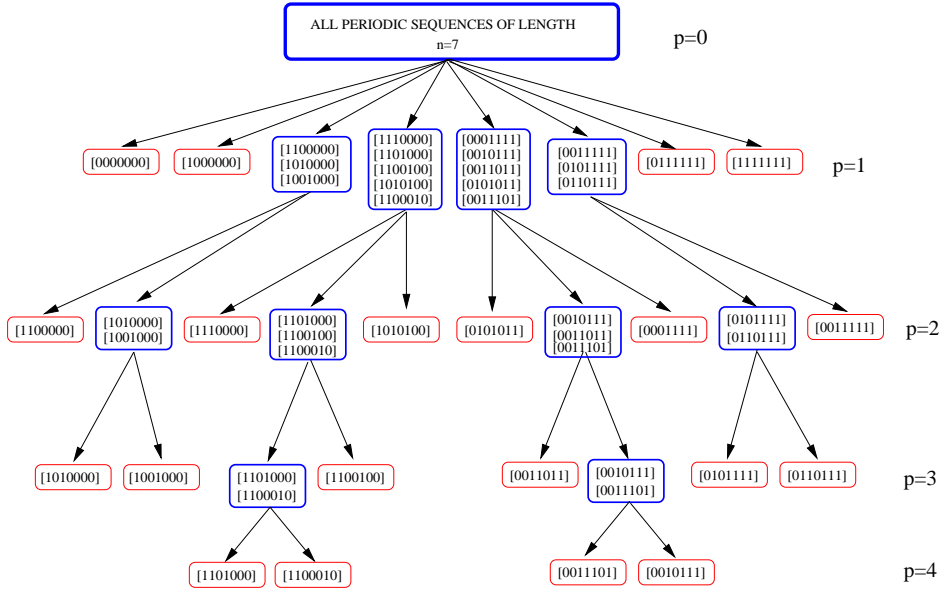}
}\end{center}
\caption{ \small{Example of hierarchy of periodic sequences for $n=7$. The ultrametric distance between two periodic sequences (shown in red boxes) is determined by the minimal cluster (blue boxes) they both belong to. E.g., $d(x,y)=4$, $d(x,z)=d(z,y)=5$ for $x=[1101000]$, $y=[1100010]$ and $z=[1100100]$. }  }\label{fig2}
\end{figure}

There are two simple but important properties of the equivalence relation $x \pclone y$ which should be emphasized:
\begin{itemize}
\item[I]
The relations $x \pclone y$ and  $x  \pclone z$ also imply that  $z \pclone y$;
\item[II]
The relation  $x \ppclone y$ implies that   $x \pclone y$.
\end{itemize}
According to the first property all periodic sequences (resp. periodic orbits) can be separated into a number of clusters $\Cll^{(p)}_i$, $i=1,\dots \N_p$  such that two sequences $x$ and $y$ (resp. $\gamma_x$, $\gamma_y$) belong to the same cluster if and only if   $x \pclone y$.  For instance, for given $p=2$ three sequences $[1101000]$, $[1100010]$, $[1100100]$ belong to the same cluster, see fig.~\ref{fig2}. In a completely analogous way one can consider clusters  $\Cl^{(p)}_i$, $i=1,\dots \N_p$ of  sequences from the set $X_n$. The connection between clusters  is given by: $\Cll^{(p)}_i= \Cl^{(p)}_i/\sim$. In other words, each $x\in \Cll^{(p)}_i$ corresponds to a set of  sequences from $\Cl^{(p)}_i$ which are related to each other by a cyclic shift.

\begin{figure}[htb]
\begin{center}{
\includegraphics[height=5.8cm]{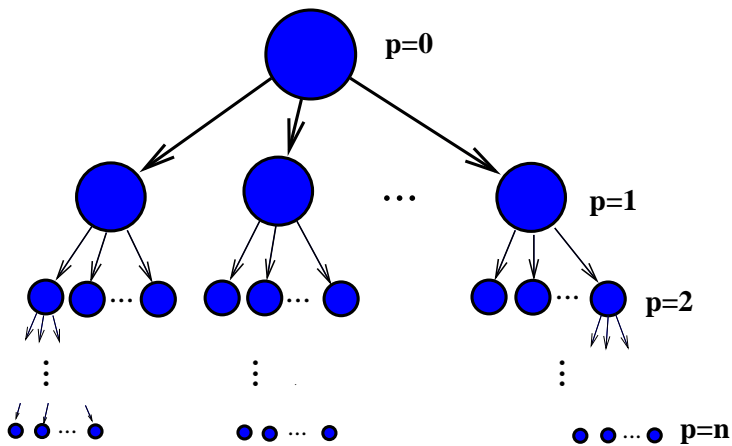}
}\end{center}
\caption{ \small{ Scheme of hierarchical structure of periodic sequences. The $p$-th level of the tree is composed of clusters (depicted by circles) of $p$-close trajectories. Each cluster at level $p$ is split into a number of clusters at the next level $p+1$.  }  }\label{fig1}
\end{figure}

The second property of $x \pclone y$ allows to organize the clusters of periodic sequences in a tree like structure. The $p$-th level of the tree contains clusters of $p$-close periodic sequences, see figs.~\ref{fig2},\ref{fig1}. One can introduce a distance in the space of sequences based on this hierarchical structure.  The distance $d(x,y)$ between two elements $x$ and $y$ being proportional to the maximal level of the tree where $x$ and  $y$ belong to the same cluster: $d(x,y)=n-\max\{ p |x  \pclone y\}$ ($d(x,x)=0$) satisfies the ultrametric property $d(x,y)\leq\max\{ d(x,z), d(z,y)\}$ \cite{VVZ1994}.  By the identification of periodic orbits $\gamma_x$, $\gamma_y$ with the corresponding sequences $x,y$, one can lift the distance $d(x,y)$ to the space of periodic orbits. And we come to the conclusion that the space $\X_n$ (or equivalently the space of periodic orbits) acquires a natural ultrametric structure.   

\subsection{Cluster distribution}

The primary  goal of the present paper is to understand how the distribution of the cluster  sizes $|\Cll^{(p)}_i|$  
 depends on the level $p$ in the limit  $n\to\infty$. 
To this end, we  need to estimate the moments  of the cluster sizes:
\begin{equation}
 \Z_k=\sum^{\N_p}_{i=1} |\Cll^{(p)}_i|^{k}. \label{mom1}
\end{equation}
It turns out, however, that a more convenient   object  to consider is 
\begin{equation}
 Z_k=\sum^{\N_p}_{i=1} |\Cl^{(p)}_i|^{k}, \label{mom2}
\end{equation}
where the sum runs over the clusters of sequences from $X_n$ (rather than sequences from $\X_n$). The connection between cluster sizes 
in  $X_n$ and $\X_n$ is particularly simple if $n$ is a prime number. In this case for each $x\in \X_n$ except $x^{(0)}=[00\dots0]$
and  $x^{(1)}=[11\dots 1]$, there are exactly $n$ sequences from  $X_n$ which are related by the cyclic shift.
Two special clusters corresponding to sequences   $x^{(0)}$, $x^{(1)}$ contain just one element. For all other clusters:
 \begin{equation}
 |\Cl^{(p)}_i|=n|\Cll^{(p)}_i|.\label{connection}
\end{equation}
This yields
\begin{equation}
 \Z_k=\frac{Z_k -2}{n^k} +2, \qquad d_n=\frac{2^n -2}{n} +2,
\end{equation}
where $d_n$ stands for the total number of elements in $\X_n$.
In the case when $n$ is not a prime number, the connection between cluster sizes  in  $X_n$ and $\X_n$ is not anymore trivial
 due to the presence of periodic orbits  with a period less than $n$. However, if one includes 
only prime periodic orbits (whose period is exactly $n$) into clusters $\Cll^{(p)}_i$, $\Cl^{(p)}_i$  then  the connection 
(\ref{connection}) remains valid. As a matter of fact, the exclusion of non-prime periodic orbits can be  justified on the ground of a
standard argument that their number is negligible  in comparison with the number of prime periodic orbits in the limit  $n\to\infty$.\footnote{The 
number of prime and non-prime periodic orbits scales as $2^n$ and $2^{n/r}$, respectively, where $r\geq 2$ is a minimal divisor of $n$ }
We arrive to the following relationship  between (\ref{mom1}) and (\ref{mom2}):
\begin{equation}
  \Z_k=\frac{Z_k}{n^k}\left(1+O(n^{-1})\right) \qquad d_n=\frac{2^n}{n}\left(1+O(n^{-1})\right). \label{superconec}
\end{equation}
Note also that for a typical cluster $|\Cl^{(p)}_i|\left(1+O(n^{-1})\right)=n|\Cll^{(p)}_i|$.

It worth of mentioning that the rescaled moments have a  simple interpretation as probabilities of finding a number   of periodic
 orbits in the same cluster. Indeed, let $\gamma_i$, $i=1,\dots k$ be a set of $k\geq 2$ orbits randomly chosen from
 the total set $\Gamma_n=\{\gamma_x| x\in \X_n\}$ of  periodic orbits of length $n$. Then the probability that all $k$ orbits belong to the same cluster is given by 
\begin{equation}
 \P_{k}=
\frac{\Z_{k}}{(d_n)^{k}}. \label{probab}
\end{equation}
In particular, the probability $ \P_{2}$ that two periodic orbits belong to the same cluster is given by $\Z_{2}/d_{n}^2$.

\subsection{Main Results}
The central result of the present paper is the following asymptotic formula for  $Z_{k}$  in the limit $n\to\infty$:
\begin{equation}
  Z_{k}
= 2^{nk}\left( \frac{1}{k}\right)^{2^{p-2}}\left(\frac{2^p}{\pi n}\right)^{(k-1)2^{p-2}}\left(1+O(n^{-1})\right).\label{moments}
\end{equation}
Using then eq.~(\ref{superconec}) and eq.~(\ref{probab}) we obtain  the probability of finding $k$ random orbits in the same cluster 
\begin{equation}
  \P_{k}
= \left( \frac{1}{k}\right)^{2^{p-2}}\left(\frac{2^p}{\pi n}\right)^{(k-1)2^{p-2}}\left(1+O(n^{-1})\right).\label{probfin}
\end{equation}

In addition, we show that  the number of periodic orbits in  the largest cluster $\Cll^{(p)}_{\max}$  is asymptotically  given by:
 \begin{equation}
 |\Cll^{(p)}_{\max}|
= \left(\frac{2^{n}}{n}\right)\left(\frac{2^p}{\pi n}\right)^{2^{p-2}}\left(1+O(n^{-1})\right).\label{maxcluster}
\end{equation}
Based on    eqs.~(\ref{moments},\ref{maxcluster}) we deduce probability  $P(t)$, $t\in [0,1]$ that  random periodic orbit from the set
 $\Gamma_n$  belongs to a cluster with   the size  less then $t |\Cll^{(p)}_{\max}|$ and show that in the limit of  $n\to\infty$ this probability depends only on  $p$:
\begin{equation}
P(t)=\int_0^t\rho(\tau)d\tau,\qquad \rho(\tau)=\frac{\left(\log \tau\right)^{2^{p-2}-1}}{(2^{p-2}-1)!}.
\end{equation}

As we show in the body of the paper the problem of counting cluster distribution of $p$-close periodic orbits is in fact equivalent to
the one of counting degeneracies in the length spectrum of the so-called {\it de Bruijn} graphs \cite{Bruijn}. In this context  the asymptotic behavior of $Z_{2}$ and related questions have been considered previously in \cite{sharp2, uzy2, Berkolajko, tanner}. The exact connection between our results and the above mentioned works is discussed in the last section of the paper.

\subsection{Organization of the paper}
The paper is organized as follows. In Sec.~3 we show that the problem of counting cluster sizes of $p$-close periodic orbits in the backer's map can be cast in the form of counting  closed paths on a certain graph passing  through the same  edges (or vertices).   Using this connection we express  $Z_k$ as a matrix  integral  of  certain type.    In Sec.~4 we evaluate these integrals in the saddle point approximation and derive  eq.~(\ref{moments}). In Sec.~5 we obtain asymptotic formula  for the size of the largest cluster $\Cll^{(p)}_{\max}$. Using then the results from  Sec.~4 we arrive  to the probability $P(t)$ of finding a periodic orbit in a cluster of the size smaller than  $t |\Cll^{(p)}_{\max}|$. Sec.~6 is devoted to the discussion of uniformity of  periodic orbits distribution over the graph.  Finally, the concluding remarks  are presented in Sec.~7.


\section{Clusters of  closed paths on  graphs}

\begin{figure}[htb]
\begin{center}{
\includegraphics[height=5.8cm]{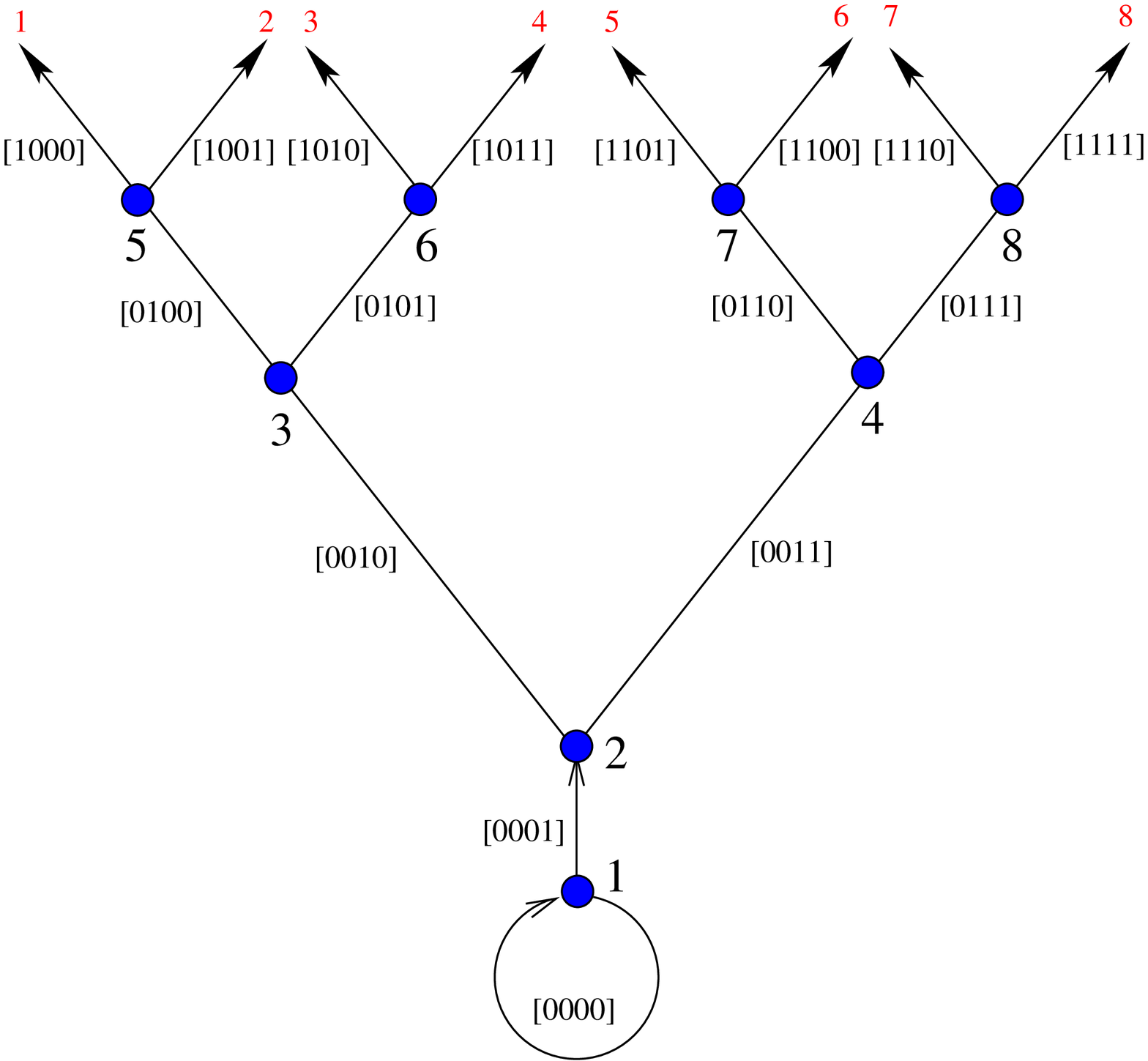} \hskip 1.2cm \includegraphics[height=5.8cm]{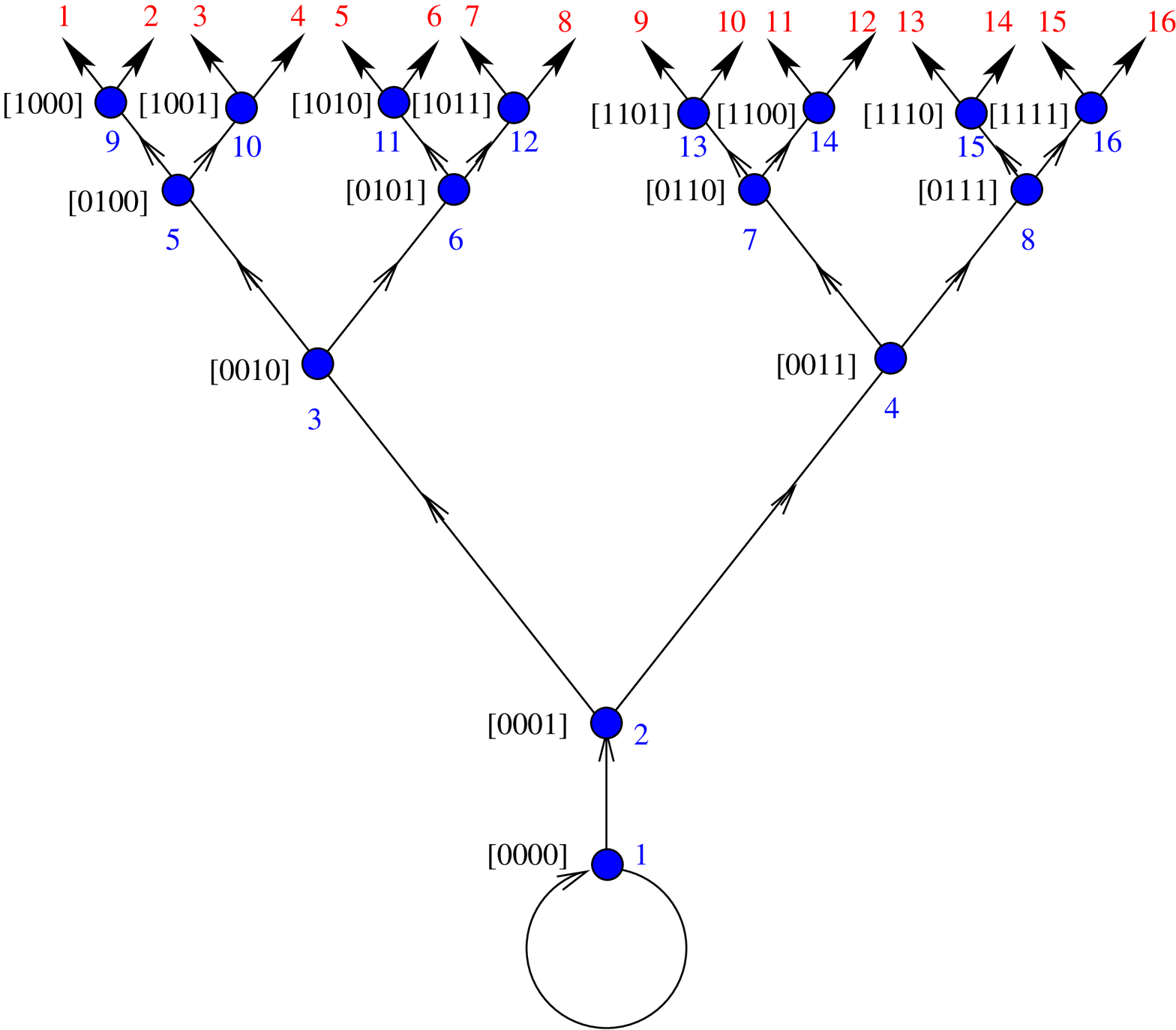}
}\end{center}
\caption{ \small{The graph $G_p$  for  $p=4$ (left) and  $p=5$ (right). Each sequence $[a_1 a_2 a_3 a_4]$, $a_i\in\{0,1\}$   encodes edges on $G_4$ and vertices on $G_5$, respectively.}  }\label{graphs}
\end{figure}

As we show below, the counting problem of   $p$-close periodic orbits is equivalent to the one of counting  closed paths on the   de Bruijn graph $G_p$  passing  the same number of times through its 
edges. The graph $G_p$ is constructed in the following way. With each sequences $a=[a_1a_2\dots a_p]$, $a\in X_p$  we associate a directed  edge $e_a$ of  $G_p$  whose    initial and terminal points are denoted by $e_a^{\scriptscriptstyle{(in)}}$ and  $e_a^{\scriptscriptstyle{(out)}}$, respectively. The connections between $2^p$  edges are fixed by the  rule: for any pair of edges $e_a$, $e_b$, defined by the sequences $a=[a_1a_2\dots a_p]$, $b=[b_1b_2\dots b_p]$ the endpoints   $e_a^{\scriptscriptstyle{(in)}}$,  $e_b^{\scriptscriptstyle{(out)}}$ belong to the same vertex if and only if $a_i=b_{i+1}$ for all $i=1,\dots p-1$, see fig.~\ref{graphs}.

 It is straightforward to see  that any closed path 
$g$  on the graph  $G_p$  passing through $n$ edges can be uniquely represented by a sequence $x=[x_1x_2\dots x_n]$ from the set $ X_n$. By such identification $i$'th edge of  $G_p$ passed by $g$  corresponds to the segment  $[x_ix_{i+1}\dots x_{p-1+i}]$ of the sequence $x$. We will use notation $g_x$ to denote closed paths corresponding to $x\in X_n$. For each closed path $g_x$  let $\bm n (x) = \{n_a, a\in X_p\}$ be the set of integers, such that $n_a$ is the number of times $g_x$ passes through the edge $a$.  Then  $x\pclone y$ if and only if   $g_x$,  $g_y$  go through every  edge of $G_p$  the same number of times (but in different time order) i.e., $\bm n (x) = \bm n (y)$.  Therefore, each cluster $\Cl_{\bm n}$ of $p$-close periodic orbits is uniquely determined by the vector of integers $\bm n=\{n_a, a\in X_p\}$.

\begin{rems} 
a) By attaching a length to each edge of the graph $G_p$ one can turn it to a metric graph. Note that each cluster $\Cl_{\bm n}$ 
consists then of trajectories having the same length. Accordingly,  counting of cluster sizes  is equivalent to counting of degeneracies in the length spectrum of the corresponding metric graph. The last problem have been studied in \cite{uzy2,sharp2,tanner,Berkolajko} for some classes of metric graphs, see also  discussion in  Sec.~7. 

b) Counting  of  closed paths   passing the same number of times through the edges of $G_p$, is actually equivalent to    counting  of closed paths passing the same number of times through the vertices of twice larger graph $G_{p+1}$. Indeed, let us  enumerate the $2^p$ vertices of the graph $G_{p+1}$ in the same way as edges of $G_p$, see fig.~\ref{graphs}. By the identification of each closed path of the length $n$ with a sequence $x\in X_n$ we obtain one-to-one correspondence between closed paths on two graphs. Correspondingly, clusters of closed paths have the same sizes.
\end{rems}

 To find the  size $|\Cl_{\bm n}|$  of $\bm n$'th cluster we need to count the number of closed paths which go through the edges $a\in X_p$ of  $G_p$  exactly $n_a$ times. To this end we introduce connectivity matrix $Q$ between edges of the graph. It is convenient to use for this purpose  a tensorial representation for the vector space on which $Q$ acts, see \cite{GO,Gu}. Let $\H$ be the $2^p$-dimensional linear space  spanned by the   vectors 
$|a\rangle= |a_1\rangle\otimes|a_2\rangle\otimes\dots\otimes|a_p\rangle$, $a_i \in\{0,1\}$. The linear operator  $Q$  acts on the vectors from  $\H$ according to the rule:
\begin{equation}\label{MatrixQ_tensor}
Q|a_1\rangle\otimes|a_2\rangle\dots\otimes|a_p\rangle=|a_2\rangle\otimes\dots\otimes|a_p\rangle\otimes\frac{1}{2}\left(|0\rangle +|1\rangle\right).
\end{equation}
Note that this definition agrees with the connectivity rules between edges on the graph $G_p$, where each edge $e_a$, $a=[a_1a_2\dots a_p]$ is connected with the  edges $e_{a'}$, $a'=[a_2\dots a_p 0]$ and   $e_{a''}$, $a''=[a_2\dots a_p 1]$.
In addition, with each edge $e_a$ we  associate a phase $\phi_a$ and define the diagonal operator  $\Lambda(\bm \phi )$, $\bm \phi:=\{\phi_a| a\in X_p\}$:
\[\Lambda(\bm \phi )|a_1\rangle\otimes|a_2\rangle\dots\otimes|a_p\rangle=e^{i\phi_a}|a_1\rangle\otimes|a_2\rangle\dots\otimes|a_p\rangle.\] 
It is straightforward to see that in the matrix form $Q$, $\Lambda(\bm \phi )$ can be written as \cite{GO}:

\begin{equation}
Q = 
\begin{pmatrix}
1&0 & \dots&0&1&0 & \dots&0\\
1&0 &\dots&0 &1&0 &\dots&0 \\
0&1 &\dots&0 &0&1 &\dots&0\\
0&1 & \dots&0 &0&1 &\dots&0\\
\vdots& \vdots& \ddots &\vdots&\vdots& \vdots& \ddots &\vdots\\
0&0 & \dots&1 & 0&0 & \dots&1\\
0&0 & \dots &1 & 0&0 & \dots &1
\end{pmatrix},
\quad
\Lambda(\bm \phi )=
\begin{pmatrix}
e^{i\phi_1}&0 &0& \dots&0 \\
0&e^{i\phi_2} &0&\dots&0  \\
\vdots& \vdots&  \ddots &\vdots& \vdots\\
0&0  &\dots&e^{i\phi_{2^p-1}} &  0\\
0&0  &\dots &0  & e^{i\phi_{2^p}}
\end{pmatrix}.
\end{equation}

   The introduction of matrices $Q$, $ \Lambda(\bm \phi )$ is useful because of the following relationship between  traces of their products and the sizes of the clusters $\Cl_{\bm n}$:
\begin{equation}
 \trace (Q \Lambda(\bm \phi ))^n=\sum_{\bm n} |\Cl_{\bm n}|\exp{\left( i(\bm n, \bm \phi )\right)}, \qquad  (\bm n, \bm \phi )=\sum_{a\in X_p}  n_a \phi_a,\quad n=\sum_{a\in X_p}  n_a,\label{KeyFormula}
\end{equation}
with the first sum running  over all clusters $\Cl_{\bm n}$. 
Eq.~(\ref{KeyFormula}) is a key component of our analysis, as it allows to express    $|\Cl_{\bm n}|$ through the traces of powers of matrix $Q \Lambda(\bm \phi )$. In particular, the second  moment $Z_2$   can be represented in the form of the integral over  $\phi_a$:
\begin{equation}
Z_2=\sum_{\bm n} |\Cl_{\bm n}|^2=\prod_{a\in X_p}\int_{0}^{2\pi}\frac{d\phi_a}{2\pi} \, |\trace (Q \Lambda(\bm \phi ))^n |^2. \label{SecondMoment}
\end{equation}
Analogously, higher order moments are given by  
\begin{equation}
Z_k=\sum_{\bm n} |\Cl_{\bm n}|^k=\prod_{j=1}^k\prod_{a\in X_p}\int_{0}^{2\pi}\frac{d\phi^{(j)}_a}{2\pi} \, \trace \left(Q \Lambda(\bm \phi^{(j)} )\right)^n\delta\left(\sum_{l=1}^k \phi^{(l)}_a\right). \label{Moments}
\end{equation}

As we show below the number of integration and dimensions of matrices in eqs.~(\ref{SecondMoment}, \ref{Moments}) can be actually  reduced by the factor of two. Note that the  $2^{p}\times 2^{p} $ matrix $Q$ can be represented as the product $Q=R S$ of the matrices
\begin{equation}\label{RS}
 R = 
\begin{pmatrix}
1&0 &0 & \dots&0\\
1&0 &0 & \dots&0 \\
0&1 &0 & \dots&0 \\
0&1 &0 & \dots&0 \\
\vdots&\vdots & \vdots& \ddots &0\\
0&0 &0 & \dots&1 \\
0&0 &0 & \dots &1
\end{pmatrix}, \qquad
S = 
\begin{pmatrix}
1& 0 &0 &0 &\dots& 1&0 &0 &0& \dots\\
0&1&0  & 0 &\dots& 0&1&0  &0& \dots\\
0& 0& 1 &0 &\dots& 0& 0& 1 &0&  \dots \\
0& 0& 0 &1 &\dots& 0& 0& 0 &1&  \dots \\
\vdots&\vdots& \vdots& \vdots&\ddots& \vdots&\vdots & \vdots& \vdots&\ddots 
\end{pmatrix},
\end{equation}
whose dimensions are $2^{p-1}\times 2^{p} $ and $2^{p}\times 2^{p-1} $, respectively.
Changing the order of the matrix product in eq.~(\ref{KeyFormula}) yields then 
\begin{equation}
 \trace (Q \Lambda(\bm \phi ))^n=\trace (S \Lambda(\bm \phi ) R)^n= \trace (Q' (\bm \phi ) )^n,
\end{equation}
where $Q' (\bm \phi )= S\Lambda(\bm \phi ) R$ is $2^{p-1}\times 2^{p-1} $ matrix of the form
\begin{equation}\label{MatrixQpsik}
Q '(\bm \phi ) =
\begin{pmatrix}
e^{i\phi_1}&0 & \dots&0&e^{i\phi_{2^{p-1}+1}}&0 & \dots&0\\
e^{i\phi_2}&0 &\dots&0 &e^{i \phi_{2^{p-1}+2}}&0 &\dots&0 \\
0&e^{i\phi_3} &\dots&0 &0&e^{i\phi_{2^{p-1}+3}} &\dots&0\\
0&e^{i\phi_4} & \dots&0 &0&e^{i \phi_{2^{p-1}+4} } &\dots&0\\
\vdots& \vdots& \ddots &0&\vdots& \vdots& \ddots &0\\
0&0 & \dots&e^{i\phi_{2^{p-1}-1}} & 0&0 & \dots&e^{i\phi_{2^{p}}}\\
0&0 & \dots &e^{i\phi_{2^{p-1}}} & 0&0 & \dots &e^{i\phi_{2^{p}}}
\end{pmatrix}.
\end{equation}
Furthermore,  using the invariance of the trace under the transformation 
\[Q' (\bm \phi )\to \Lambda(\bm \delta )Q' (\bm \phi )\Lambda^{-1}(\bm \delta ),\]
where $\Lambda(\bm \delta )$ is an arbitrary diagonal matrix, we can exclude half of the integration variables in  eqs.~(\ref{SecondMoment}, \ref{Moments}), see Appendix~A.  The result is the following expression for the moments 
\begin{equation}
Z_k=\prod_{j=1}^k\prod_{i=1}^{2^{p-1}}\int_{0}^{2\pi}\frac{d\phi^{(j)}_i}{2\pi} \, \trace \left(\Q (\bm \phi^{(j)} )\right)^n\delta\left(\sum_{l=1}^k \phi^{(l)}_i\right), \label{FinalMoments}
\end{equation}
with the matrix $\Q (\bm \phi)$  given by
\begin{equation}\label{FinalMatrix}
\Q (\bm \phi ) = 
\begin{pmatrix}
e^{i\phi_{1}}&0 & \dots&0&1&0 & \dots&0\\
e^{i \phi_{2}}&0 &\dots&0 &1&0 &\dots&0 \\
0&e^{i\phi_{3}} &\dots&0 &0&1 &\dots&0\\
0&e^{i \phi_{4} } & \dots&0 &0&1&\dots&0\\
\vdots& \vdots& \ddots &0&\vdots& \vdots& \ddots &0\\
0&0 & \dots&e^{i\phi_{2^{p-1}-1}} & 0&0 & \dots&1\\
0&0 & \dots &e^{i\phi_{2^{p-1}}} & 0&0 & \dots &1
\end{pmatrix}.
\end{equation}
In the next section we will use eq.~(\ref{FinalMoments})  to obtain large $n$ asymptotic of $Z_k$.

\section{Clustering Probability of  $k$ orbits }

In order to evaluate moments $Z_k$ we will  apply saddle point approximation to the integral (\ref{FinalMoments}), where $n$ will play the role of a large parameter. To make the exposition more transparent  we first consider below the second moment and later extend the result to all $k>2$. 

\subsection{Saddle point approximation for $Z_2$ } For $k=2$ the integral (\ref{FinalMoments}) can be written as

\begin{equation}
Z_2=\left(\prod_{i=1}^{2^{p-1}}\int_{0}^{2\pi}\frac{d\phi_i}{2\pi}\right)\cdot \exp{\F_n(\bm \phi )},  \qquad  \F_n(\bm \phi )=
\log|\trace (\Q(\bm \phi ))^n |^2. \label{SecondMoment2}
\end{equation}
It is easy to see that the global maximum of $\F_n(\bm \phi )$ is attained at $\bm \phi =\bm0$, where   $\phi_i =0$, $i=1,\dots 2^{p-1}$ and $\F_n(\bm 0)=n\log 2$. We therefore need 
 to expand  $\F_n(\bm \phi )$  around zero up to the second order in $\bm \phi$ and then  use saddle point approximation in (\ref{SecondMoment2}). To evaluate derivatives of $\F_n(\bm \phi )$ it is  convenient to use the following decomposition of the  matrix  $\Q(\bm \phi )$:
\[ \Q(\bm \phi )=\Lambda(\bm \phi )Q_0+Q_1.\]
Here $Q_0=[R\; 0]$, $Q_1=[0\; R]$ are  matrices composed of two blocks. The first (resp. second) block is given by the matrix $R$ (from eq.~(\ref{RS})) of the dimension $2^{p-1}\times 2^{p-2}$, while the second (resp. first) one is  $2^{p-1}\times 2^{p-2}$ matrix of zeroes (for the element-wise definition of $Q_0,Q_1$, see Appendix~B). Straightforward calculations give then:
\begin{equation}
 \frac{\partial \F_n(\bm \phi )}{\partial\phi_j}= 2n\Re\left[\frac{i\trace\left(\Lambda(\bm \phi )P_j Q_0 (\Q(\bm \phi ))^{n-1}\right)}{\trace(\Q(\bm \phi ))^{n}}\right],
\label{FirstDeriv}
\end{equation}
where $P_j$ denotes projection matrix on $j$ element of the basis, i.e., $(P_j)_{m,l}=\delta_{m,i}\delta_{l,i}$. It follows immediately 
that  $\frac{\partial \Q(\bm \phi )}{\partial\phi_j}\rvert_{\bm \phi=0}=0$, as it should be for a saddle point. Taking an additional derivative   
in (\ref{FirstDeriv}) yields
\begin{multline}
\frac{\partial^2 \F_n(\bm \phi )}{\partial\phi_j \partial\phi_i}\Big{\rvert}_{\bm \phi =0}=2n
\left[ 
\frac{-\sum_{k=0}^{n-2}\trace\left(P_i Q_0  Q^{k} P_j Q_0 Q^{n-k-2}
\right) +\delta_{i,j}\trace\left( P_i Q_0 Q^{n-1} \right) }{\trace \, Q^{n}} \right]  \\
+2n^2\left[ \frac{\trace\left( P_iQ_0 Q^{n-1}\right) \trace\left( P_jQ_0 Q^{n-1}\right) }{ (\trace \, Q^{n})^2  } 
\right] ,
\label{SecondtDeriv}
\end{multline}
implying that
\begin{equation}
\frac{\partial^2 \F_n(\bm \phi )}{\partial\phi_j \partial\phi_i}\Big{\rvert}_{\bm \phi =0}=-n B_{i,j},
\end{equation}
where the matrix $B$ is defined by:
\begin{equation}
B=2^{-p-1}\left(\bar{Q}+\bar{Q}^T +2 I -(1+2p)\left(\frac{Q}{2}\right)^p \right), \qquad \bar{Q}= Q_0\sum_{k=0}^{p-1}
\left(\frac{Q}{2}\right)^k.\label{MatrixB}
\end{equation}
We can use now (\ref{SecondMoment2}) to  evaluate $Z_2$ in the large $n$ limit:
\begin{align}\label{MomentZ2_result}
 Z_2& =2^{2n}\left(\prod_{j=1}^{2^{p-1}}\int_{0}^{2\pi}\frac{d\phi_j}{2\pi} \right)\cdot\exp{\left[-\frac{n}{2}\sum_{i,j} B_{i,j}\phi_i \phi_j\right]}\left(1+O\left(\frac{1}{n}\right) \right)\nonumber\\
& =2^{2n}(2\pi n)^{-2^{p-2}} \left(\det B\right)^{\frac{1}{2}} \left(1+O\left(\frac{1}{n}\right) \right). 
\end{align}
The determinant of $B$ can be explicitly calculated (see Appendix~B) which finally leads to:
\begin{equation}
 Z_2 (n)=2^{2n}\left(\frac{2^{p-1}}{\pi n}\right)^{2^{p-2}} \left(1+O\left(\frac{1}{n}\right) \right). \label{FinalSecondMoment}
\end{equation}

\subsection{Saddle point approximation for $Z_k$, $k>2$ }  Our starting 
point is the  representation (\ref{FinalMoments}) for  $Z_k$:
\begin{gather}
Z_k=\left(\prod_{j=1}^{k-1}\prod_{i}^{2^{p-1}}\int_{0}^{2\pi}\frac{d\phi^{(j)}_i}{2\pi}\right) \cdot \exp{\F_n(\{\bm \phi^{(j)}\} )},\label{FinalMoments2}\\
\F^{(k)}_n(\{\bm \phi^{(j)}\} )=\log\left[\trace \left(\Q \left(- \bar{\bm \phi}  \right)\right)^n \prod_{i=1}^{k-1} \trace \left(\Q \left(\bm \phi^{(i)} \right)\right)^n
\right], \nonumber
\end{gather}
where we introduced notation:
\[ \bar{\bm \phi}:= \sum_{j=1}^{k-1}\bm \phi^{(j)}.\]
As in the case $k=2$, the maximum of $\F^{(k)}_n$ is attained when all phases  vanish i.e., for $\bm \phi^{(j)}=\bm 0, j=1,\dots k-1$.
After taking the first derivative of $\F^{(k)}_n$ we have 
\begin{multline}
 \frac{\partial \F^{(k)}_n(\bm \phi )}{\partial\phi^{(l)}_j}=  n\left[\frac{i\trace\left(\Lambda(\bm \phi^{(l)} )P_j Q_0 (\Q(\bm \phi^{(l)} ))^{n-1}\right)}{\trace(\Q(\bm \phi^{(l)} ))^{n}} \right]\\
- n\left[\frac{i\trace\left(\Lambda(-\bar{\bm \phi}  )P_j Q_0 (\Q(- \bar{\bm \phi} ))^{n-1}\right)}{\trace(\Q(- \bar{\bm \phi} ))^{n}}
\right].
\label{FirstDeriv1}
\end{multline}
The second derivative of $\F^{(k)}_n$ at    $\bm \phi^{(j)}=0, j=1,\dots k-1$ is then:
\begin{equation}
\frac{\partial^2 \F^{(k)}_n(\bm \phi )}{\partial\phi^{(l)}_j \partial\phi_i^{(m)}}\Big{\rvert}_{\bm \phi =0}:=-n B^{(k)}_{[i, l;j, m]},
\end{equation}
with  the   elements of the  $2^{p(k-1)}\times 2^{p(k-1)}$  matrix $B^{(k)}$ given by:
\[B^{(k)}_{[i, l;j, m]}=\frac{1}{2} B_{i,j}(\delta_{l,m}+1).\]
Note that  $B^{(k)}$ can also be written as the product of two matrices composed of $(k-1)\times(k-1)$ blocks: 
\begin{equation}
B^{(k)}=
\frac{1}{2}\begin{pmatrix}
 2\1&  \1&\dots & \1\\
 \1& 2\1&\dots &\1\\
\vdots& \vdots& \ddots& \vdots\\
\1& \1 &\dots & 2\1
\end{pmatrix}
\begin{pmatrix}
B& \0 &\dots & \0 \\
 \0 & B &\dots & \0 \\
\vdots& \vdots& \ddots& \vdots\\
\0 & \0 &\dots & B
\end{pmatrix},\label{MatrixForm}
\end{equation}
with $\1$ and   $\0$ standing for unit and zero  $2^{p}\times 2^{p}$ matrices, respectively.  

Applying now saddle point approximation to (\ref{FinalMoments2}) we obtain
\begin{multline}
 Z_k =2^{kn}\left(\prod_{l=1}^{k-1}\prod_{j=1}^{2^{p-1}}\int_{0}^{2\pi}\frac{d\phi^{(l)}_j}{2\pi} \right)\cdot\exp{\left[-\frac{n}{2}\sum_{i,j} B^{(k)}_{[i, l;j, m]}\phi^{(l)}_i \phi^{(m)}_j\right]}\left(1+O\left(\frac{1}{n}\right) \right)\\
 =2^{kn}(2\pi n)^{-(k-1)2^{p-2}} \left(\det B^{(k)}\right)^{\frac{1}{2}} \left(1+O\left(\frac{1}{n}\right) \right).  \label{AlmostFinal}
\end{multline}
By eq.~(\ref{MatrixForm}) the determinant of $B^{(k)}$ can be explicitly evaluated:
\[\det B^{(k)}=\left(\frac{k}{2^{k-1}}\right)^{2^{p}}\left(\det B\right)^{k-1}.\]
Substituting this expression into eq.~(\ref{AlmostFinal}) gives then
\begin{equation}
 Z_k=\frac{2^{2k}}{k^{2^{p-2}}}\left(\frac{2^{p}}{\pi n}\right)^{(k-1)2^{p-2}} \left(1+O\left(\frac{1}{n}\right) \right).\label{Final_k_Moment}
\end{equation}

\begin{figure}[htb]
\begin{center}{
\includegraphics[height=5.8cm]{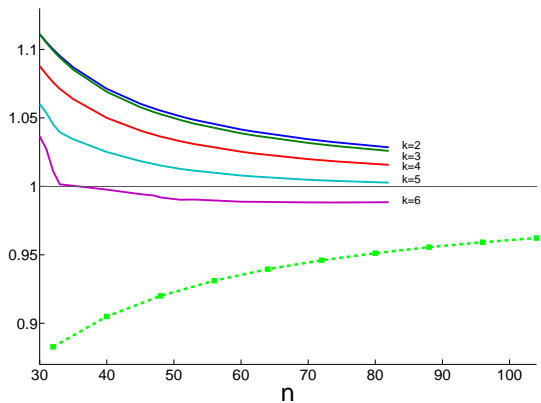} 
}\end{center}
\caption{ \small{The upper five curves show the ratio of the exact moments $Z_k$, $k=2,\dots 5$  to the  asymptotics  (\ref{Final_k_Moment}) as a function of $n$ in the case $p=3$. They are  ordered with respect to $k$ with the most  upper curve corresponding to $k=2$. The dashed curve at the bottom of the plot is  the ratio of the exact size of  the largest cluster  to its asymptotic value (\ref{maxcluster}) measured at the points $n$ which are multiplies of eight. }  }\label{figmoments}
\end{figure}

To verify the above asymptotic formula we  evaluated 
the ratio between the leading order term of (\ref{Final_k_Moment}) and the exact value of $Z_k$ obtained by numerical calculations of $|\Cl_{\bm n}|$. The results  for $p=3$ are presented on fig.~(\ref{figmoments}). As can be  observed,  this ratio is close to one for  large $n$.

\section{Distribution of cluster sizes}

We will use now the results of the previous section to evaluate the probability $P(t)$ that a randomly chosen periodic orbit belongs to 
a cluster of the size $t|\Cl_{\max}|$, $t\in [0,1]$, where $|\Cl_{\max}|=\max_{\bm n}|\Cl_{\bm n}|$ is the size of the largest cluster. To this end we first need to establish the asymptotic behavior of  $|\Cl_{\max}|$.

 By eq.~(\ref{KeyFormula}) the maximal size is given by:  
\begin{gather}
 |\Cl_{\max}|=\max_{\bm n}\left( \prod_{j=1}^{2^{p-1}} \int_{0}^{2\pi} \frac{d\phi_j}{2\pi}  \right) \cdot\exp{\F(\bm \phi )},\\
\F(\bm \phi )=-i({\bm n},\bm \phi )+\log\trace\left( \Q\Lambda(\bm \phi )\right)\nonumber.
\end{gather}
To evaluate the above expression we can  once again use saddle point approximation. Taking the first derivative of $\F(\bm \phi )$ 
leads to the following equation for the saddle point
\begin{equation}
{n}_i=n\frac{\trace\left(\Lambda(\bm \phi )P_i Q_0 (\Q(\bm \phi ))^{n-1}\right)}{\trace(\Q(\bm \phi ))^{n}},
\label{FirstDeriv3}
\end{equation}
with ${n}_i$ being $i$'s component of the vector $\bm n$.
It can be expected that the cluster of the  maximum size   is provided  by the most homogeneous vector $\bar{\bm n}$ among all admissible vectors $\bm n$. For the sake of simplicity  take $n$, such that  $n\mod 2^p=0$.   In this case    $\bar{\bm n}=2^{-p}(1,1,\dots 1)$ and the saddle point equation (\ref{FirstDeriv3}) is satisfied when  $\bm \phi=\bm 0$. At this point   $\F(\bm 0 )=n\log 2$. Since $|\F(\bm \phi )|\leq n\log 2$ for any $\bm \phi$, it is clear that $\Cl_{\bar{\bm n}}$ is, indeed, has (at least asymptotically) the largest  size among all clusters.  Repeating then the same calculations for the second derivative of $\F(\bm \phi )$, as in the previous section,  we obtain after saddle point approximation:
\begin{equation}
|\Cl_{\max}|=Z_2\left({n}/{2} \right)\left(1+O\left( {1}/{n} \right)\right),\label{Finalmaxcluster}
\end{equation}
where $Z_2(n)$ is given by eq.~(\ref{FinalSecondMoment}).

To calculate the probability $P(t)$, it is useful to notice that the moments $Z_k$ can be represented as integrals:
\begin{equation}
 \frac{Z_k}{2^n|\Cl_{\max}|^{k-1}}=\int_{0}^{1}dt \rho(t) t^{k-1}, \label{MomRepresentation}
\end{equation}
where $\rho(t) \Delta t$ is the probability to find a random periodic orbit in a cluster $\Cl_{\bm n}$ whose size belongs to the interval $|\Cl_{\max}|\cdot t\leq|\Cl_{\bm n}|\leq|\Cl_{\max}|\cdot(t+ \Delta t)$. The probability to find a random periodic orbit in the cluster  of the size smaller than $|\Cl_{\max}|t$ is therefore: 
\[ P(t)=\int_{0}^{t}d\tau\rho(\tau).\]

\begin{figure}[htb]
\begin{center}{
\includegraphics[height=5.8cm]{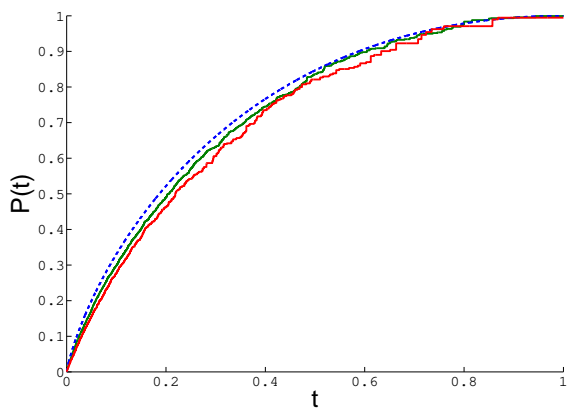} 
}\end{center}
\caption{ \small{ The exact distribution of cluster sizes for $n=70$ (upper green) and $n=47$ (lower red) is shown in comparison with the asymptotic expression  $P(t)=t (\log t -1)$  (dashed blue line) in the case $p=3$}  }\label{distribution}
\end{figure}

It remains  to find  $\rho(\tau)$. After substituting into eq.~(\ref{MomRepresentation}) the asymptotic expressions for $Z_k$ and $|\Cl_{\max}|$  we obtain  in the limit $n\to\infty$:
\begin{equation}
 \int_{0}^{1}dt \rho(t) t^{k-1}=k^{-2^{p-2}}.\label{EqForRho}
\end{equation}
By taking the Laplace transform on both sides of eq.~(\ref{EqForRho}) one has  
 \begin{equation}
 \rho(t)=\frac{\left(\log t\right)^{2^{p-2}-1}}{(2^{p-2}-1)!}.
\end{equation}
In particular, for $p=2,3$ this gives:
\begin{equation}
 P(t)=t \quad \mbox{  (for $p=2$) }; \qquad P(t)=t (\log t -1) \quad  \mbox{  (for $p=3$) }.
\end{equation}
The comparison of the above result with the direct numerical simulation is shown on fig.~\ref{distribution}.

\section{Anisotropy  properties of clusters} 
The method used in the previous sections to calculate moments $Z_k$ can be  also applied to obtain a more refined  information
on    the distribution of periodic orbits in clusters with regard to each edge individually. One can ask for instance how many times
a sequence $a\in X_p$ appears in a random sequence $x$ of length $n$. As one can expect, if all orbits are weighted in the same way
the result does not depend on $a$:
\begin{equation}
<n_a>:=2^{-n}\sum_{\bm n} n_a |\Cl_{\bm n} |=-2^{-n}i\partial_{\phi_a}\trace\left( Q\Lambda(\bm \phi )\right)^n|_{\bm\phi =0}=n/2^p. \label{anys1}
\end{equation}
In other words, the periodic orbits are ergodically distributed over the graph $G_p$. However,  if only  periodic orbits from large clusters (or small clusters) are considered the question regarding homogeneity of their distribution over the edges of the graph, does not seem to have a trivial answer. For instance it is clear that smallest clusters are dominated by  periodic orbits with either large number of zeroes or ones. Therefore, among the periodic orbits belonging to small clusters there should be enhanced probability to meet
a subsequence consisting of all zeroes (or ones). 

\begin{figure}[htb]
\begin{center}{
\includegraphics[height=4.5cm]{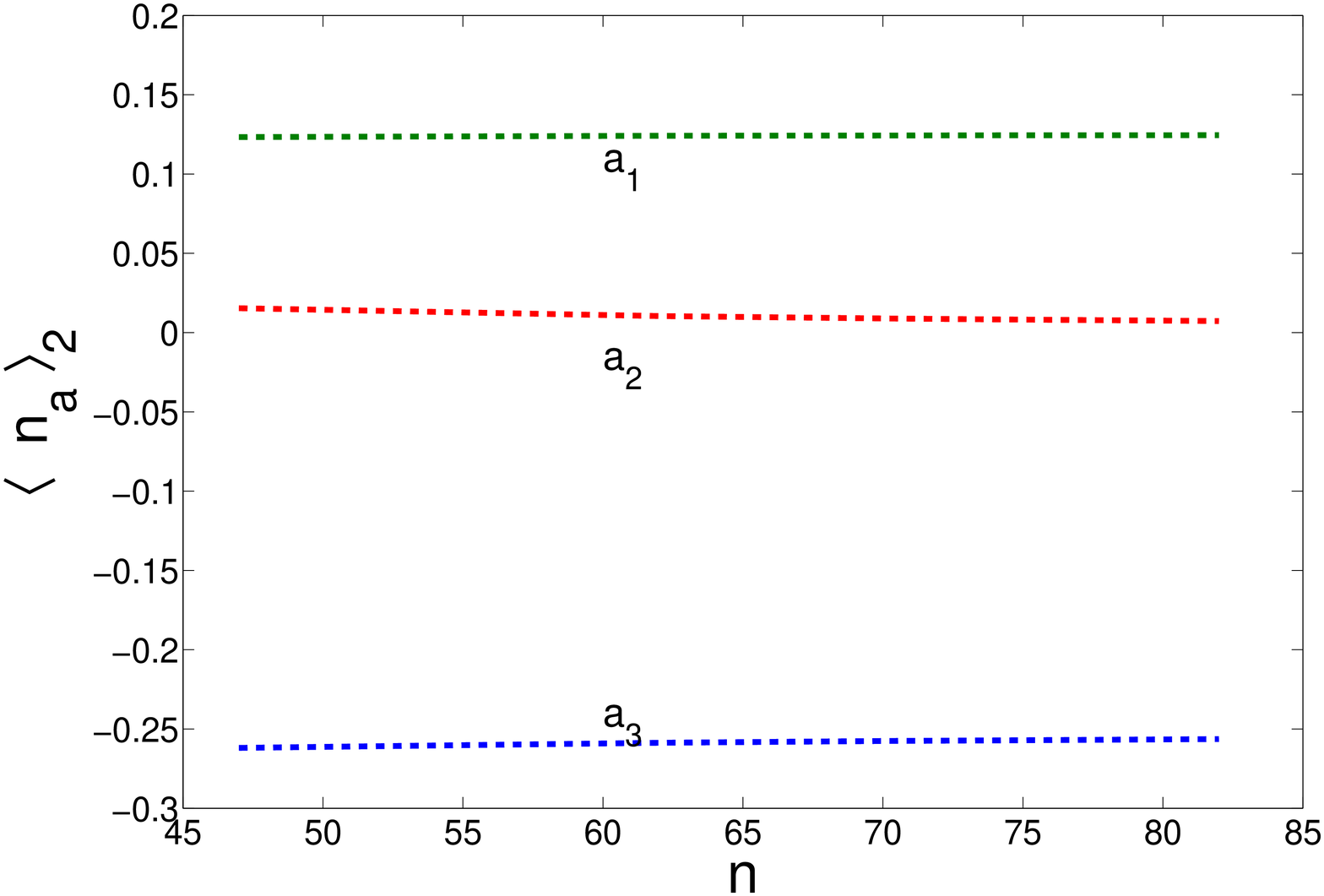}\hskip 9.1cm 
\includegraphics[height=4.5cm,width=4cm]{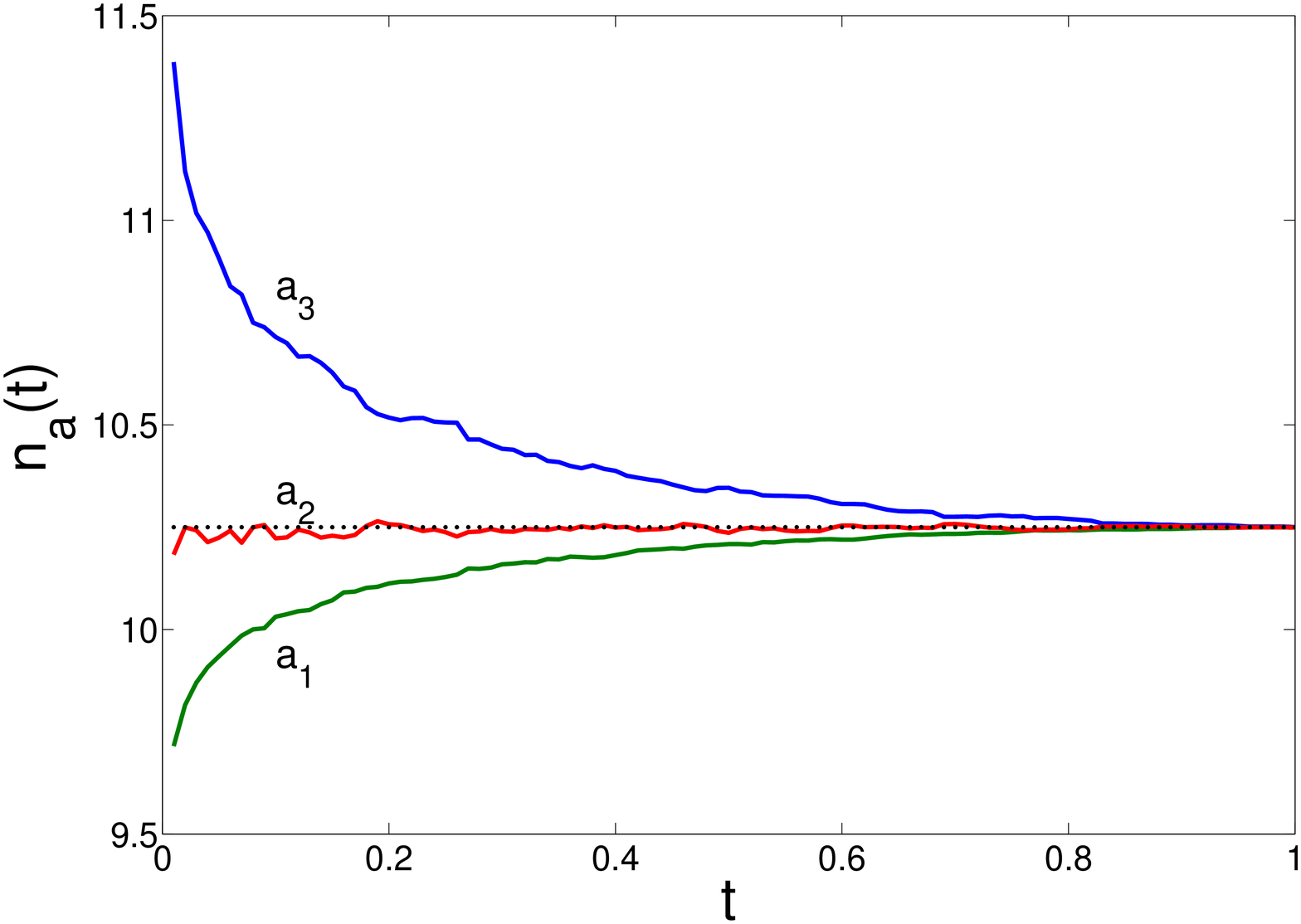}\hskip 0.1cm 
\includegraphics[height=4.5cm]{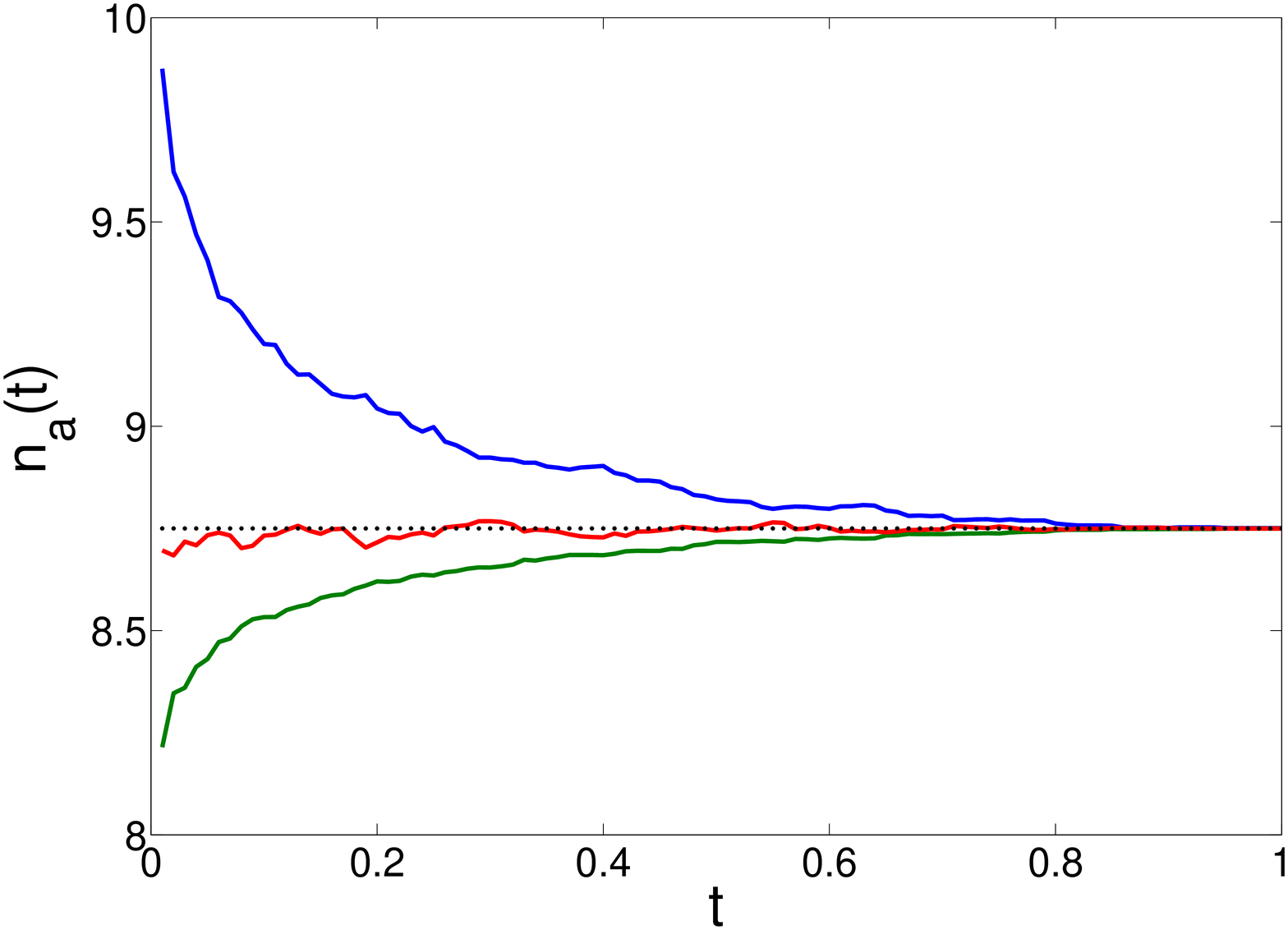}
}\end{center}
\caption{ \small{ Distribution of periodic orbits over edges of the graph for $p=3$. The  upper figure shows the graph of $<n_a>_2$   as a function of $n$ for three different edges $a_1=[011]$, $a_2=[101]$ and $a_3=[000]$.  The lower figures show the graphs (for the same edges) of $\bar{n}_a(t)$  for $n=82$ (left)  and $n=70$ (right), respectively. Here the doted lines correspond to the constant value $n/8$. Note, that up to a shift in the vertical direction, these two graphs are  close to each other.}  }\label{figanysotropy}
\end{figure}

The purpose of the present section is to investigate the anisotropy properties of the graph with regard to distribution of periodic orbits belonging to clusters of different size. To this end let us consider the sum
\begin{equation}
<n_a>_2:=\frac{\sum_{\bm n} n_a |\Cl_{\bm n} |^2}{\sum_{\bm n}  |\Cl_{\bm n} |^2}. \label{anys2}
\end{equation}
As opposed to $<n_a>$ the above quantity gives a larger weight to the periodic orbits from larger clusters. Thus, in principle,    
$<n_a>_2$ might depend on $a$. To verify this, we observe that  (\ref{anys2}) can be written as 
\begin{equation}
<n_a>_2 :=\frac{1}{Z_2}\left(\prod_{j=1}^{2^{p-1}}\int_{0}^{2\pi}\frac{d\phi_j}{2\pi}\right)\cdot \exp{\F^{(a)}_n(\bm \phi )},\label{AnysSecondMoment}
\end{equation} 
\begin{multline}
  \F^{(a)}_n(\bm \phi )=
\log\left[\Re(-i\partial_{\phi_a}\trace (\Q(\bm \phi ))^n\trace (\Q(-\bm \phi ))^n )\right]\nonumber\\
=\log \left[n\Re(\langle a|\Lambda(\bm \phi ) Q_0 Q(\bm \phi )^{n-1}|a\rangle\trace (\Q(-\bm \phi ))^n )\right]
, \nonumber
\end{multline}
where $|a\rangle$ is the state corresponding to the sequence (edge) $a$.
We can now apply saddle point approximation to  eq.~(\ref{AnysSecondMoment}). Expanding $\F_n(\bm \phi )$ up to the second order in 
$\phi_j$'s gives 
\begin{equation}
 \F^{(a)}_n(\bm \phi )=\log(n2^{2n-p})-\frac{n}{2}\left(\sum_{i,j}B_{i,j}\phi_i\phi_j +O(\phi_i^4)\right), \label{expansion}
\end{equation}
with the matrix $B$ given by eq.~(\ref{MatrixB}). Substituting then  (\ref{expansion}) into (\ref{AnysSecondMoment}) leads  to 
\begin{equation}
<n_a>_2=n/2^p+O(n^0),
\end{equation}
where only the subleading term (possibly) depends on $a$. This form of $<n_a>_2$ indicates that to the leading order in $n$ the periodic orbits are equidistributed uniformly over the graph. The asymmetry shows up only in the second and higher order terms of the  asymptotic expansion. In fig.~\ref{figanysotropy} we plotted    $<n_a>_2$ as a function of $n$ for different edges of the graph. It is clearly visible that existing asymmetry between different edges, essentially, does not  grow with  $n$. 

In an  analogous way one can estimate   averages:
\begin{equation}
<n_a>_k :=\frac{\sum_{\bm n} n_a |\Cl_{\bm n} |^k}{\sum_{\bm n}  |\Cl_{\bm n} |^k}. \label{anysk}
\end{equation}
for an arbitrary $k$. As in the case of $k=2$,  the leading order term of the asymptotic expansion  of $<n_a>_k$ is equal to $n/2^p$. This can be interpreted to the extent that   periodic orbits  uniformly (to the leading order of $n$) pass through the edges of the graph $G_p$, independently of the size of the clusters they belong.  The non-uniformity, however, does appear in the next order $n^0$ of the asymptotic expansion. To demonstrate this, we  calculate numerically  the averages:
\begin{equation}
\bar{n}_a(t):=\frac{\sum\limits_{|\Cl_{\bm n} |< t|\Cl_{\max}|} n_a |\Cl_{\bm n} |}{\sum\limits_{|\Cl_{\bm n} |< t|\Cl_{\max}|} |\Cl_{\bm n} |}, \label{anys4}
\end{equation}
where the sum runs over clusters with  sizes less than $t|\Cl_{\max}|$, $t\in [0,1]$. It follows straightforwardly from the definition of $ \bar{n}_a(t)$  and  eq.~(\ref{anys1}) that
\[ \sum_{a\in X_p} \bar{n}_a(t)=n \mbox{ for  $t\in [0,1]$},\qquad \bar{n}_a(1)=n/2^p.\]
 In the case $p=3$, the resulting plot of    $\bar{n}_a(t)$  is shown on fig.~(\ref{figanysotropy}) for $n=82$ and  $n=60$, respectively. As one can see, there is  enhanced probability for periodic orbits from small clusters $t\ll 1$ to pass through the edge $a=[000]$, and suppressed probability to pass through the edge $[011]$. On the other hand this picture depends very little on $n$.

\section{Discussion}

To summarize, the problem of counting $p$-close periodic orbits of the backer's map can be cast in an equivalent form of finding degeneracies in the length spectrum of the de Bruijn graph $G_p$. The latter  problem has previously attracted attention in \cite{sharp2}, where an asymptotic expression for $\Z_2$ has been obtained for a generic graph. Related  counting problems were also  considered in \cite{uzy2} for fully connected graphs and in \cite{tanner} for binary directed graphs. In comparison to \cite{sharp2}  we go somewhat  further, as we derive asymptotics for all moments $\Z_k$ and   explicitly obtain the leading term prefactors depending on $p$.  Our considerations  have been restricted to a simplest possible symbolic dynamics which occurs in the backer's map (leading to a specific binary graph). 
As a matter of fact, the  results can be straightforwardly generalized to the  alphabets  with a larger number  of symbols.   Moreover, the present approach can be  extended to the symbolic dynamics  with  non-trivial grammar rules, when certain symbolic subsequences are forbidden.  In that case the resulting graph $G_p$ would typically  have a non-homogeneous structure, where the number of outgoing and  incoming edges depend on a specific vertex. 

The rescaled moments $\Z_k$ can be interpreted  as the probabilities $\P_k$  that $k$ randomly chosen periodic sequences are  $p$-close. Making use of information on $\P_k$, we obtain the asymptotics of the probability  $P(t)$ to find  a random periodic sequence in a cluster of the size smaller than $t|\Cll_{\max}|$, $t\leq 1$, with $\Cll_{\max}$ being the largest cluster. Most significantly,  in the large $n$ limit,   $P(t)$  does not depend on the length of sequence, but only on $p$.
It is worth noting that  $P(t)$ is basically determined   by the distribution of ``large`` clusters, whose size is of the order $|\Cll_{\max}|$. 
``Small`` clusters $\Cll_i$, whose size is $|\Cll_i|\ll |\Cll_{\max}|$  do not affect (asymptotically) the moments  $\Z_k$ and therefore, do not contribute to $P(t)$. A natural question arises: what is the number of ``large`` clusters in comparison to the number of ``small'' clusters? 
As we show in Appendix~C  the total number of clusters $\N_p$ is proportional to $n^{2^{p-1}}$. On the other hand, based on eqs.~(\ref{Finalmaxcluster}, \ref{Final_k_Moment}) a rough estimation of ``large`` clusters yields 
\[ \N_p^{(l)}\sim \Z_k/|\Cll_{\max}|^k \sim n^{2^{p-2}}.\]
This shows that the number of ``large`` clusters scales as $\N_p^{(l)}\sim\sqrt{\N_p}$. In other words most of the  clusters are ``small'' in the limit $n\to\infty$, although their influence on $\Z_k$ is negligible.
 
  We have also studied the distribution  of  the periodic orbits over the  graph $G_p$. A priori, it might be expected that subsequences consisting of only  zeroes (or ones) appear more often than others  in periodic sequences  belonging to ``small`` clusters. As we show, this is indeed so, but  such inhomogeneities vanish  asymptotically.  To the leading order of $n$ periodic orbits cover the graph $G_p$ uniformly. In other words,  in  long sequences belonging to clusters of size comparable with  $|\Cll_{\max}|$  each subsequence of $p$ symbols occurs on average the same number of times.

 One of the main 
 motivations for the present study comes from  the theory of 
 quantum chaos, where clusters of $p$-close orbits play an important  role. It is worth  mentioning, however, that the limit 
considered in the present paper, $n\to \infty$ with a fixed $p$, is somewhat different from the usual semiclassical limit. In the last case spectral correlations are determined by pairs of long periodic orbits whose encounter lengths are  proportional to the logarithm of their total lengths  \cite{haake1}.    This corresponds to the limit of both  $n\to\infty$, $p\to\infty$ such that the ratio $n/2^p$ is fixed.
 In this limit  one might expect completely different behavior for $\P_k$. In particular the number of $p$-close pairs should grow
faster than $2^n$.  The exact behavior of  $\P_k$ in the semiclassical limit is of great interest and we leave it for future investigations.

\section*{Acknowledgments}
We thank  P. Braun and M. Kieburg  for valuable discussions. This work was supported by the Sonderforschungsbereich Transregio 12

\appendix

\section{Change of integration variables in~(\ref{SecondMoment}) and~(\ref{Moments}).} \label{Appendix1}
Here we demonstrate how the total number of integration variables in the integral~(\ref{SecondMoment})  can be effectively reduced by half. To perform the change of variables we use the invariance of the trace under the multiplication of each matrix $Q\Lambda(\bm\phi)$ by a diagonal matrix $\Lambda(\bm \xi)=\diag{e^{i\xi_1},\dots,e^{i\xi_{2^p} }} $ and its conjugate from the left and from the right side correspondingly:
$$
Q'\Lambda(\bm\phi)\to  e^{-i\xi}\Lambda(\bm \xi) Q'\Lambda(\bm\phi)\Lambda(\bm \xi)^\dag.
$$
Any such transformation with an arbitrary $\bm \xi=(\xi_1,\dots \xi_{2^{p-1}})$ and $\xi$ leaves the integrand in (\ref{SecondMoment}, \ref{Moments}) intact. Note also that  this transformation does not affect the structure of the matrix (\ref{MatrixQpsik}), but only its phases.  Using this one can eliminate the phases  ${\phi}'_{1}\equiv\phi_{2^{p-1}+1}, \dots {\phi}'_{2^{p-1}}\equiv\phi_{2^{p}} $ on the ``right side'' of the matrix $Q'$  by imposing the system of linear equations on $\xi_k$, $k=1,\dots,2^{p-1}$ 
\begin{eqnarray}\label{Variable_Change1}
\bar{\phi}_{2k-1}-\xi_{2k-1}+\xi_{2^{p-1}+k}-\xi&=&0;\\
\bar{\phi}_{2k}-\xi_{2k}+\xi_{2^{p-1}+k}-\xi&=&0.\nonumber
\end{eqnarray}
On the ``left side'' of the matrix $Q'$ the above transformation induces new phases
\begin{eqnarray}\label{Variable_Change2}
\varphi_{2k-1}&=&\phi_{2k-1}-\xi_{2k-1}+\xi_k-\xi;\\
\varphi_{2k}&=&\phi_{2k}-\xi_{2k}+\xi_k-\xi,\nonumber
\end{eqnarray}
where $\xi_{i}$'s are determined by  eq.~(\ref{Variable_Change1}).
After fixing the common phase $\xi$ to be a symmetric combination of other variables $\xi_k$, i.e. $\xi=2^{-p}\sum_k\xi_k$ the system of equations (\ref{Variable_Change1}-\ref{Variable_Change2}) can be cast into the matrix form:
\begin{eqnarray}\label{Variable_Change3}
{\bm\phi'}=F\bm\xi;\\
\bm\varphi=\bm\phi-G \bm\xi,\label{Variable_Change4}
\end{eqnarray}
where the matrices $F$ and $G$ are expressed in terms of $Q_0$, $Q_1$, and $Q$: 
\begin{equation}\label{Matrices_F_and_G}
G=\1-Q_0+(Q/2)^p  ;\qquad F=\1 -Q_1+(Q/2)^p .
\end{equation}

Combining now  eq.~(\ref{Variable_Change3}) with eq.~(\ref{Variable_Change4}) we define the following linear transformation
\begin{equation}\label{Jacobian}
\left(\begin{array}{c}
\bm\phi\\
\bm{\phi}'
\end{array}\right)=
\left(\begin{array}{cc}
\1&GF^{-1}\\
0&\1
\end{array}\right)
\left(\begin{array}{c}
\bm\varphi\\
{\bm\varphi}'
\end{array}\right),
\end{equation}
which exists since the determinants of both $F$ and $G$ are different from zero (see proposition~\ref{Proposition1}).
By construction, the resulting matrix $e^{-i\xi}\Lambda(\bm \xi) Q'\Lambda(\bm\phi)\Lambda(\bm \xi)^\dag$ has the required form (\ref{FinalMatrix}) with the phases given by $\bm \varphi$.  Furthermore, since the Jacobian of the linear transformation (\ref{Jacobian})  equals  one, the  variables ${\bm \varphi}'$ do not enter the integrand at all and can be integrated out.
 In the case of higher moments $Z_k$, $k>2$ one has to perform the similar transform for each set of variables $\bm\phi^{(j)}$. The resulting expression for the integral is given then by the equations~(\ref{FinalMoments}) and~(\ref{FinalMatrix}).

\section{Calculation of the determinant of the  matrix (\ref{MatrixB})}\label{Appendix2}
In this appendix we calculate the determinant of the matrix $B$.
We recall that $B$ is defined using  $2^p\times 2^p$ matrices  $Q_0$, $Q_1$, their sum $Q=Q_0+Q_1$ and $Q_p=\left(\frac{Q}{2}\right)^p$, see (\ref{MatrixB}). For the sake of completeness, we give their   element-wise definition: 
\begin{itemize}
\item The entries of the matrices $Q$ and $Q_0$, $Q_1$ are given by ($i,j =1,\dots 2^p$):
\begin{gather}\label{MatrixQ_definition}
[Q]_{i\;j}=\sum_{k=1}^{2^{p-1}}(\delta_{i,2k-1}+\delta_{i,2k})(\delta_{j,k}+\delta_{j,2^{p-1}+k})\\
 [Q_0]_{i\;j}=\sum_{k=1}^{2^{p-1}}(\delta_{i,2k-1}+\delta_{i,2k})\delta_{j,k};\quad
{}[Q_1]_{i\;j}=\sum_{k=1}^{2^{p-1}}(\delta_{i,2k-1}+\delta_{i,2k})\delta_{j,2^{p-1}+k}.
\end{gather}

\item All elements of matrix $Q_p$  are equal  to $2^{-p}$. 
\end{itemize}

Bellow we collect a number of  matrix relations which are of use in the further analysis. They follow directly from the matrix definitions. For any natural~$k$:
\begin{equation}\label{matrix_relations}
Q^kQ_p=Q_pQ^k=2^kQ_p;\qquad Q_0^kQ_p=Q_1^kQ_p=Q_p;\qquad Q_p^k=Q_p;
\end{equation}
\begin{equation}\label{Q1_Q0_strange_relation}
Q_0^TQ_0+Q_1^TQ_1=2\cdot\1.
\end{equation}
 For  the traces of various matrix products we have:
\begin{equation}\label{matrixQ0_trace}
\trace Q_p Q_{0,1}^k=\trace Q_{0, 1}^k=1, \qquad \trace Q^k=2^k,
\end{equation}
where (and further) $Q_{0,1}$ stands either for $Q_0$ or $Q_1$.

Two more, useful equalities can be derived by using~(\ref{matrix_relations})
\begin{eqnarray}\label{Matrices_in_power1}
(Q_{0,1}-Q_p)^k&=&(Q_{0,1}-Q_p)Q_{0,1}^{k-1};\\
(Q-Q_p)^k&=&Q^k-(2^k-1)Q_p.\label{Matrices_in_power2}
\end{eqnarray}
Both can be proved by induction. To derive~(\ref{Matrices_in_power1}) it is enough to notice that $Q_p(Q_{0,1}-Q_p)=0$. To obtain the coefficient at $Q_p$ in eq.~(\ref{Matrices_in_power2}) we assume that $(Q-Q_p)^k=Q^k-a_kQ_p$ with some yet unknown $a_k$. Then the next iteration gives the following recurrence relation for $a_k$: $a_{k+1}=2^k+a_k$. Its solution is $a_k=2^k-1$ as it is indicated in~(\ref{Matrices_in_power2}).

Having described the properties of the matrices $Q_{0,1}$, $Q$ and $Q_p$ we come to calculation of their determinants. The first axillary statement is: 
{\prop For any real $\alpha$
\begin{equation}\det\left[\1-\alpha(Q_{0,1}-Q_p)\right]= 1, \qquad \det \left[\1-\alpha({Q}-{Q_p})\right]=1-\alpha. \label{Useful_determinant}\end{equation}
 \label{Proposition1}} 

\begin{proof}
Consider the logarithm of the determinant:
$$
\log\det[\1-\alpha(Q_{0,1}-Q_p)]= -\sum_{k=1}^\infty \frac{\alpha^k}{k}\trace(Q_0-Q_p)^k.
$$
Applying now the formula~(\ref{Matrices_in_power1}) we come to the matrix $(Q_0-Q_p)Q_0^k$ which trace is zero for all $k>0$ due to the relations~(\ref{matrixQ0_trace}). This proves the first statement.
In the same way, using the equation~(\ref{Matrices_in_power2}) and the information about traces we derive
$$
\log\det \left[\1-\alpha({Q}-{Q_p})\right] = - \sum_{k=1}^\infty \frac{\alpha^k}{k}\trace\left(Q-Q_p\right)^k=\log(1- \alpha).
$$
\end{proof}
We can use   the above proposition for  $\alpha=1$ to obtain the  determinants of the matrices $F$ and $G$ defined in~(\ref{Matrices_F_and_G}):
{\col 
\[\det F=\det G=1\]. \label{Corollary 1}} 
\\

In the remaining part of the appendix the determinant of $2^p\times 2^p$ matrix~(\ref{MatrixB})
$$
B=2^{-p-1}\left(2\1 -(1+2p)Q_p+Q_0 \sum_{r=0}^{p-1}\left(\frac{Q}{2}\right)^r+\sum_{r=0}^{p-1} \left(\frac{Q^T}{2}\right)^rQ_0^T  \right),
$$ is evaluated. The following remark helps to perform the calculations.
{\rem In the saddle-point calculations of the integral~(\ref{FinalMoments})  made after the change of variables one comes to the Gaussian integral~(\ref{MomentZ2_result}) involving matrix $B$. However, change of the order leads to another matrix, $2^{-p}M_{p+1}$ . It has twice larger dimensionality and connected to the matrix $B$ by the relation
\begin{multline}\label{Matrices_connection}
2^{-p}M_{p+1}=
\left(\begin{array}{cc}
\1&0\\
-GF^{-1}&\1
\end{array}\right)
\left(\begin{array}{cc}
B&0\\
0&0
\end{array}\right)\left(\begin{array}{cc}
\1&-GF^{-1}\\
0&\1
\end{array}\right)\\=\left(\begin{array}{cc}
B&-BGF^{-1}\\
-\big(GF^{-1}\big)^TB&\big(GF^{-1}\big)^TBGF^{-1}
\end{array}\right).
\end{multline}
Explicit form of matrix $M_{p+1}$ is
\begin{equation}\label{Matrix_M}
M_{p+1}=\1+\sum_{r=1}^p\left[  \left(\frac{Q}{2}\right)^r+  \left(\frac{Q^T}{2}\right)^r\right]- (2p+1) Q_{p+1},
\end{equation} 
where all matrices are of the size $2^{p+1}\times 2^{p+1}$.
}

Eq.~(\ref{Matrices_connection}) allows to cast  the problem of calculation of $ \det B$ into a ``symmetric'' form. Namely, by the following proposition 
the determinant of $B$ can be expressed through the eigenvalues of two matrices which depend only  on $Q$ and $Q^T$.
{\prop The determinant of the matrix $B$ can be represented as
\begin{equation}\label{DetB}
\det B=2^{-2^p}\frac{4\prod_{\lambda_j\ne0}\lambda_j}{\det\tilde{M}_p},
\end{equation}
where $\lambda_j$ are non-zero eigenvalues of $2^{p+1}\times 2^{p+1}$ matrix $M_{p+1}$ and $\tilde{M}_p$ is $2^p\times 2^p$ matrix $\tilde{M}_p:=\1+M_p+3 Q_p$. 
}

 \begin{proof}
Consider the spectral problem for the matrix $M_{p+1}$:
$$
M_{p+1}\left(
		\begin{array}{c}
			\bm X\\
			\bm Y
		\end{array}
	\right)=\lambda \left(
		\begin{array}{c}
			\bm X\\
			\bm Y
		\end{array}
	\right)\Longrightarrow \begin{array}{l}
			 B \bm X- BC \bm Y=\lambda \bm X\\
			-C^T B \bm X + C^TB C\bm Y=\lambda \bm Y,
	\end{array}
$$ 
where $C=GF^{-1}$.
The latter system reduces to the equation 
$$
\lambda(C^T\bm X+  \bm Y)=0,
$$
meaning that either $\lambda=0$ or  $-C^T\bm X=\bm Y$. Backward substitution results in the following eigenvalue problem
$$
B(\1+C C^T)\bm X=\lambda \bm X
$$
where $\lambda$'s are non-zero eigenvalues of the matrix $M$, so that we can write down
$$
\det B=\frac{\prod_{\lambda_j\ne0}\lambda_j}{\det(\1+ C C^T)}=\frac{\prod_{\lambda_j\ne0}\lambda_j}{\det(F^T F+G^T G)}.
$$
In the last expression the result of the proposition~\ref{Proposition1} has been used.

One can proceed further by evolving the denominator. First we observe that it depends only on the matrix $Q$. Indeed, due to the properties~(\ref{matrix_relations}),~(\ref{Q1_Q0_strange_relation}) one has
\begin{multline*}
F^T F+G^T G=4\cdot\1-Q-Q^T+2Q_p\\=2\left(\1-\frac{Q}{2}+\frac{Q_p}{2}\right)\left(\frac{1}{\1-\frac{Q}{2}+\frac{Q_p}{2}}+\frac{1}{\1-\frac{Q^T}{2}+\frac{Q_p}{2}}\right)\left(\1-\frac{Q^T}{2}+\frac{Q_p}{2}\right)
\end{multline*} 
Expansion of each of the fractions with the help of equation~(\ref{Matrices_in_power2}),
$$
\frac{1}{\1-\frac{Q}{2}+\frac{Q_p}{2}}=
\sum_{k=0}\frac{(Q-Q_p)^k}{2^k}=\sum_{k=0}^{p-1}\left(\frac{Q}{2}\right)^k-(p-2)Q_p,
$$
yields
\begin{equation}
F^T F+G^T G
=2\left(\1-\frac{Q}{2}+\frac{Q_p}{2}\right)\left(\1+M_p+3 Q_p\right)\left(\1-\frac{Q^T}{2}+\frac{Q_p}{2}\right),
\end{equation} 
where all matrices are of the size $2^p\times 2^p$. By taking into account~(\ref{Useful_determinant}) we arrive to the final expression for the determinant of $B$.
\end{proof}

It remains now to calculate the spectrum of matrices $M_{p+1}$ and $\tilde{M}_{p}$. Note that both matrices have a similar structure and can be treated in the same way. Their spectra is  provided by  the following  lemma.
{\lemma \label{lemma1}The spectrum of the matrix $M_{p+1}$  contains $2^p$ (half of the total number) zero eigenvalues, the non-zero part of the spectrum has the following structure:
 $$
 \lambda_k=k+1\quad \mbox{ with multiplicity } \quad 2^{p-k-1},\qquad 0\le k \le p-1,
 $$
 and $\lambda_p=p+1$ has multiplicity one.   The non-trivial eigenvalues  of $\tilde{M}_{p}$ are given by:
$$
 \tilde\lambda_k=k+2\quad \mbox{ with multiplicity } \quad 2^{p-k-2},\qquad 0 \le  k \le p-2,
 $$
and $\tilde\lambda_{p-1}=p+1$,  $\tilde\lambda_p=4$ having  multiplicity one. The remaining  $2^{p-1}-1$ eigenvalues are all equal to one.
  }
\begin{proof}
To prove this lemma it is convenient to work within the tensorial representation of matrices. According to~(\ref{MatrixQ_tensor}) the matrix $Q^r$ is given by 
\begin{equation}\label{Operator_Q}
Q^r=\left(\underbrace{s\otimes\dots \otimes s}_{r\;\; \mbox{\scriptsize times}}\otimes \1\otimes\dots\otimes \1\right)\cdot T^r, 
\end{equation}
where $T$ is the following  shift operator:
$$
T|a_1\rangle\otimes|a_2\rangle\dots\otimes|a_p\rangle=|a_2\rangle\otimes\dots\otimes|a_p\rangle\otimes|a_1\rangle,
$$ 
and the projection  $s$ acts on $|0\rangle,|1\rangle$ as
\[
s|0\rangle=0;\qquad
s|1\rangle=|1\rangle.\]
Note that the result of the action of the operator $Q^r$ on the basis vectors  $|j_1\rangle\otimes|j_2\rangle\otimes\dots\otimes|j_p\rangle$, $j_k\in\{0,1\}$ essentially  depends on the number of consecutive ``ones" at the end of the  sequence $j_1j_2\dots j_p$. Whenever  this  number is larger than $r$, the result of the action of $Q^r$ is zero. This property allows to find  all eigenvalues of the operator $M_{p+1}$. 

Let $\chi^{(k)}$ be a vector  having exactly $k$ ``ones'' at the right end of the encoding sequence, separated by two ``zeroes'' (if $k< p$) from the rest of the sequence , i.e.,
$$
\chi^{(k)}=|0\rangle\otimes\omega_{p-k-1}\otimes|0\rangle\otimes\underbrace{|1\rangle\otimes\dots\otimes|1\rangle}_{k\;\; \mbox{\scriptsize times}},
$$
where $\omega_{p-k-1}=|j_1\rangle\otimes|j_2\rangle\otimes\dots\otimes|j_{p-k-1}\rangle$ is an arbitrary product vector of the length $p-k-1$. 
By the definition   $\chi^{(p+1)}$  is the vector consisting of only ``ones''. It is easy to see that $\chi^{(p+1)}$ is an eigenvector of $M_{p+1}$ with zero eigenvalue.  
All other eigenvectors of $M_{p+1}$ can be constructed as  linear combinations of the rotations of $\chi^{(k)}$'s:
\begin{equation}\label{eigenvecors_of_operator_M}
\sum_{j=0}^k\alpha_j\bm T^j\chi^{(k)}.
\end{equation}
The action of $M_{p+1}$ on each of these combinations ($k\le p$) results in
$$
\bm M_{p+1}\sum_{j=0}^k\alpha_j\bm T^j\chi^{(k)}=\left(\sum_{j=0}^k\alpha_j\right)\left(\sum_{j'=0}^k\bm T^{j'}\chi^{(k)}\right).
$$
Equating the right hand side of this expression with $\lambda\sum_{j=0}^k\alpha_j\bm T^j\chi^{(k)}$ we rewrite the original problem as the  eigenvalue problem for the $k\times k$ matrix consisting of all ones:
$$
\sum_{j=0}^k\alpha_j=\lambda \alpha_k,\qquad k=0,1,\dots,p.
$$
The  solution is well known -- all eigenvalues are zeros except  one which is equal to $k+1$. The degeneracy of the eigenvalues is defined by the free part $\omega_{p-k-1}$ of the vector $\chi^{(k)}$  and, therefore, equals to $2^{p-k-1}$ for $0\le k\le p-1$ and  to one for $k=p$. The total number of non-zero eigenvalues is
$$
1+\sum_{k=0}^{p-1}2^{p-k-1}=2^p.
$$

The eigenvalues of $\tilde{M}_p$ can be obtained by a general shift of all eigenvalues of $M_{p}$ by $1$ with the only exception of zero eigenvalue corresponding to $\chi^{(p)}$ which should be shifted by $4$. 
\end{proof} 
 
 By Lemma~\ref{lemma1} one has the following chain of identities
 \begin{equation}
 \prod_{\lambda_j\ne0}\lambda_j=(p+1)\prod_{k=0}^{p-1}(k+1)^{2^{p-k-1}}
=(p+1)\prod_{k=0}^{p-2}(k+2)^{2^{p-k-2}}=\frac{1}{4}\det\tilde{M}_p. 
 \end{equation}
Substituting  this  into~(\ref{DetB}) gives the final expression for the determinant of $B$:
{\prop
$$
\det B=2^{-2^{p}}.
$$
 }

\section{Calculation of the total number of clusters}\label{Appendix3}
Here we estimate the total number $\mathcal{N}_p(n)$ of equivalence classes
$\Cl_{\bm n}$ of all sequences of the total length $n$ generated by the
equivalence relation $x\pclone y$ ($x,y\in \X_n$). Recall that every equivalence class
is uniquely parametrized  by a vector of integers $\bm n=\set{n_a, a\in
X_p}$. The elements of this vector determine the number of times  a periodic orbit from   $\Cl_{\bm n}$ passes trough the
corresponding edge of the graph $G_p$.   Since vector $\bm n$ corresponds to a real periodic orbit of the length $n$, its components must satisfy the  following constraints:

\begin{itemize}
\item[i.] The total length of the trajectory is fixed:
\begin{equation}\label{Currents_conservation1}
\sum_{a\in X_p}n_a=n.
\end{equation}
\item[ii.]
The number of times a periodic orbit enters a vertex of the graph $G_p$ must be equal to the number of exits from the same vertex.
This balancing  condition is represented by the equation
\begin{equation}\label{Currents_conservation}
S\bm n=\bm n^T R.
\end{equation}
\end{itemize}
It is easy to see that for any vector $\bm n$ satisfying the above conditions,  with $n_a\neq 0$ for all $a$, it is possible to find a closed path on the graph which passes through an edge $a$ exactly  $n_a$ times. Note also that the number of  solutions   with $n_a= 0$ for some $a$ is smaller by a factor $1/n$ than the total number  of solutions of (\ref{Currents_conservation1},\ref{Currents_conservation}), see \cite{Berkolajko}. Therefore, to find the leading asymptotics of  $\mathcal{N}_p(n)$ it is sufficient to count vectors of  positive integers satisfying the equations~(\ref{Currents_conservation1},\ref{Currents_conservation}). 
  
Since the system~(\ref{Currents_conservation}) is composed of
 $2^{p-1}-1$ linearly independent conditions, we can chose $2^{p-1}$ first elements of $\bm n$ freely, while the rest is then uniquely fixed by  eqs.~(\ref{Currents_conservation1},\ref{Currents_conservation}). In addition, the  constraints $n_i\geq 0$, $i=1,\dots2^{p}$ must be satisfied.  These constraints  define a  $2^{p-1}$-polytope $\mathcal{V}_p$ in the $2^{p-1}$-dimensional space of  $n_1,\dots n_{2^{p-1}}$.
Geometrically the number of clusters
$\mathcal{N}_p(n)$ can be interpreted as the total number of  points with integer coordinates 
 encompassed by  $\mathcal{V}_p$.  Accordingly, the  leading term of  $\mathcal{N}_p(n)$ in the
large-$n$ limit  is given by the volume of $\mathcal{V}_p$. Therefore,
  \begin{equation}
\mathcal{N}_p(n)={w}_p n^{2^{p-1}}(1+O(1/n)), \label{c3}
  \end{equation}
where the coefficient ${w}_p$ can be calculated explicitly for low values of $p$. We illustrate this by the following example.

\noindent{\bf Example:} For $p=2$ the  conditions~(\ref{Currents_conservation1},\ref{Currents_conservation}) take  the form:
\[ n_2=n_3, \qquad n_1 +n_2 +n_3+n_4=n.\]
We chose two independent integer be $n_1=k$, $n_2=m$. Since $n_i\geq 0$ for all $i=1,2,3, 4$, the problem is reduced to  calculation of the area of the triangle
\[\mathcal{V}_2=\{x\geq 0, y\geq 0, n-2x-y\geq 0\},\]
in the $(x,y)$-plane.
As a result, for  $p=2$ we obtain $\mathcal{N}_2(n)= n^2/4 + O(n)$.

It worth mentioning that similar problem for non-directed graphs was  considered in \cite{Berkolajko}. It was shown that   the number of  equivalence classes (equiv. the degeneracy classes in the length  spectrum of the graph) in the leading order of $n$  is proportional to $n^{|E|-1}$, where $|E|$ is the total number of edges in the graph. For  comparison with (\ref{c3}), note that  the number of edges in $G_p$ is $2^p$. This  reflects the fact  that the number of  equivalence classes in directed graphs is essentially smaller than in non-directed graphs.

\end{document}